\documentclass[11pt]{amsart}
\usepackage[margin=1in]{geometry}
\usepackage{graphicx}
\usepackage{subfig}
\usepackage{comment}
\usepackage{amstext} % for \text macro
\usepackage{array}   % for \newcolumntype macro
\newcolumntype{C}{>{$}c<{$}} % math-mode version of "l" column type
\usepackage[justification=centering]{caption}
\usepackage{color}
\newcommand{\gr}[1]{{\color{black} #1}}

\begin{document}
\title[Modified MME for Generalized Laplace]{Modified Method of Moments\\ for Generalized Laplace Distribution}
\author{Adrian Fischer, Robert E. Gaunt, Andrey Sarantsev}

\address{D\'epartement de Math\'ematique, Universit\'e Libre de Bruxelles}

\email{Adrian.Fischer@ulb.be}

\address{Department of Mathematics, The University of Manchester}

\email{robert.gaunt@manchester.ac.uk}

\address{Department of Mathematics \& Statistics, University of Nevada, Reno}

\email{asarantsev@unr.edu}

\begin{abstract}
In this note, we consider the performance of the classic method of moments for parameter estimation of symmetric variance-gamma (generalized Laplace) distributions. We do this through both theoretical analysis (multivariate delta method) and a comprehensive simulation study with comparison to maximum likelihood estimation, finding performance is often unsatisfactory. In addition, we modify the method of moments by taking absolute moments to improve efficiency; in particular, our simulation studies demonstrate that our modified estimators have significantly improved performance for parameter values typically encountered in financial modelling, and is also competitive with maximum likelihood estimation.  

\vspace{3mm}

\noindent{{\bf{Keywords:}}} Variance-gamma distribution; method of moments; parameter estimation
%In this note, we consider the performance of the classic method of moments for parameter estimation of symmetric variance-gamma (generalized Laplace) distributions. Through both theoretical analysis (multivariate delta method) and a comprehensive simulation study, we demonstrate that caution must be used in applying the method of moments for generalized asymmetric Laplace distributions and related models, as performance is often unsatisfactory. In addition, we modify the method of moments by taking absolute moments to improve efficiency; in particular, our simulation studies demonstrate that our modified estimators have significantly improved performance for parameter values typically encountered in financial modelling. 
\end{abstract}

\maketitle

\theoremstyle{plain}
\newtheorem{thm}{Theorem}
\newtheorem{lemma}[thm]{Lemma}
\newtheorem{prop}[thm]{Proposition}
\newtheorem{cor}[thm]{Corollary}

\theoremstyle{definition}
\newtheorem{defn}{Definition}
\newtheorem{asmp}{Assumption}

\theoremstyle{remark}
\newtheorem{rmk}{Remark}
\newtheorem{exm}{Example}

\thispagestyle{empty}

\section{Introduction}

\subsection{Generalized asymmetric Laplace (variance-gamma) distributions} 

In recent decades, a family of {\it generalized asymmetric Laplace} (\textsc{GAL}), or {\it variance-gamma} (\textsc{VG}) distributions has gathered the attention of researchers. A random variable $X$ with distribution from this family has {\it characteristic function} (Fouier transform)
\begin{equation}
\label{eq:FT-VG}
\mathbb E[e^{i\omega X}] = e^{im\omega}\cdot\left[1 + ic\omega + b\omega^2/2\right]^{-a}.
\end{equation}
Here, $a, b, c, m$ are real-valued parameters, with restrictions $a, b > 0$. 
%Their interpretation: $a$ is {\it shape}, $b$ is {\it scale}, $c$ is {\it skew}, $m$ is {\it location}. 
The probability density function (PDF) is known and involves the modified Bessel function of the second kind; see equation (\ref{svgdefn}) for an expression in the case $c=0$. This family was introduced (for the case $c=0$) into the financial literature in a seminal work of \cite{VG-Market} under the name {\it variance-gamma} and later independently in \cite{GAL1} under the name {\it generalized asymmetric Laplace}; see a more detailed exposition of a multivariate version in \cite{GAL2}. The class of {\it asymmetric Laplace} distributions was studied in the book \cite{Book}, which also mentions \textsc{GAL}; see also \cite{vgsurvey} for an up to date review. This random variable has moments of all orders, and finite {\it moment generating function} (\textsc{MGF}) $\mathbb E[e^{tX}]$ for $t$ in a neighborhood of zero. Its tails are heavier than Gaussian. It can be represented as a {\it mean-variance Gaussian mixture:} 
\begin{equation}
\label{eq:mixture}
X|G \sim \mathcal N(m-cG, bG),\quad G \sim \Gamma(a, 1),
\end{equation}
where the density of the $\Gamma(a, 1)$ distribution is $f_a(x) = \frac{x^{a-1}}{\Gamma(a)}e^{-x}$, $x>0$. In addition, the GAL distribution is {\it infinitely divisible:} For each $n = 2, 3, \ldots$ we can represent $X = X_1 + \ldots + X_n$ (on a certain probability space), where $X_1, \ldots, X_n$ are independent and identically distributed. Moreover, each $X_i$ has a distribution which belongs to the same family, with changed parameters. Thus we can create a L\'evy process $L(\cdot)$ with increments distributed as GAL. This is called {\it Laplace motion}, \cite{LaplaceMotion}. From~\eqref{eq:mixture}, we can derive that this process can be represented as a Brownian motion $W(\cdot)$ (with non-trivial drift and diffusion) subordinated by a gamma process $\Gamma(\cdot)$: 
%(a nondecreasing L\'evy process with increments having gamma distributions): 
$L(t) = W(\Gamma(t))$. These properties make this family valuable for financial modeling, as in \cite{Levy}; see applications to option pricing in \cite{VG-Option} and fitting financial data in \cite{VG}.  

%A generalization of this family are the {\it generalized Normal-Laplace} (\textsc{GNL}) random variables, which can be represented as a sum of two independent random variables: $Y = X + Z$, with $X$ \textsc{GAL} and $Z \sim \mathcal N(0, \sigma^2)$. This  is a 5-parameter family of distributions, introduced in \cite{Reed1}. Many important properties of the \textsc{GAL} distribution: finite moments, heavier-than-Gaussian tails, infinite divisibility, remain true. The corresponding L\'evy process with increments having these  distributions is a sum of two independent processes: Laplace and Brownian motions, see \cite{Reed2}. We can similarly represent \textsc{GNL} variables as a mean-variance Gaussian mixture, as in~\eqref{eq:mixture}. 

%Another generalization of the variance-gamma family is {\it generalized hyperbolic} (GH) distributions, introduced by Ole Barndorff-Nielsen in \cite{GH1, GH2}; see also the bibliography in \cite{GH}. This is another five-parameter family of distributions, widely used in financial econometrics. However, it is not infinitely divisible and therefore does not have an associated L\'evy process.

\subsection{Estimation methods} Parameter estimation, however, remains difficult for the variance-gamma distribution. A direct maximum likelihood estimation (\textsc{MLE}) is computationally difficult because of  the presence of the modified Bessel function of the second kind in the density formula for \textsc{GAL}; see various methods in the thesis \cite{Thesis} and a similar approach to autoregressive models in \cite{ECM}. Representation as in~\eqref{eq:mixture} opens the door for expectation-maximization algorithms (\textsc{EM}). 
%Characteristic functions as in~\eqref{eq:FT-VG} have a relatively simple form. One could try to use them to estimate. But this was not yet done. 
The thesis \cite{Thesis} contains several estimation methods which are modifications of MLE. 

\subsection{Method of moments estimation} In some literature, perhaps the simplest method was used: method of moments estimation (\textsc{MME}). For example, we compute the first 4 moments of the variance-gamma distribution. We solve the resulting system of equations explicitly. This gives us an expression of parameters via moments. Finally, we substitute empirical moments in place of exact ones. This gives us parameter estimators. See \cite{NVM} for MME and generalizations for \textsc{VG} distributions.
%, \cite[Section 2.1]{Reed2} for \textsc{GNL}, and \cite{GNL-AR} for autoregression of order $1$ with innovations distributed as \textsc{GNL}. 
In addition, \cite{VG} obtains MME for skewness parameter $c$ approaching zero, and removing terms of order $c^2, c^3$ and higher from expressions of moments. 

It is straightforward to show that these estimators $\hat{\bf{\theta}}$ are consistent (converge to the true values $\bf{\theta}$ as the sample size $N$ tends to infinity) and asymptotically normal: $N^{1/2}(\hat{\bf \theta} - {\bf \theta}) \to_d \mathcal N(0, \Sigma)$ as $N \to \infty$, where $\Sigma$ is the limiting covariance matrix. However, these estimators are not efficient: The variances in this limiting matrix are larger than the variances for the \textsc{MLE}. 

In several articles, the \textsc{MME} is mentioned for estimation of these distributions and related time series models. This gives an impression that \textsc{MME} is acceptable for parameter estimation for these families. However, this question has typically been addressed with applications to financial modelling in mind with quite specialized parameter values considered.
%However, this question was not studied thoroughly and rigorously in the existing literature. 
To the best of our knowledge, no rigorous simulation or theoretical study has been conducted to study actual applicability and to quantify errors for this method across the full range of parameter values that may appear in applications. This note fills this gap \gr{in the symmetric $c=0$ case}. 

\subsection{Our contributions} In this article, we focus on the symmetric variance-gamma (generalized Laplace) distribution, where $c = 0$ in~\eqref{eq:mixture}. Through theoretical results and a simulation study, with comparison to the classic MLE, we test the performance of MME, finding performance is often unsatisfactory.
%We provide a simulation study which shows that errors in \textsc{MME} are often too large to use this method. Also, we get theoretical results: asymptotic normality, with covariance matrix too large for meaningful inference. 
%This shows that implicit assumptions in the existing literature about applicability of the delta method are false. 
%We have every reason to believe that if this is true for the symmetric variance-gamma case, this is even more so for more general case of \textsc{GAL} with unknown $c\in\mathbb{R}$. 

Next, we modify \textsc{MME} to improve efficiency. Specifically, the location parameter $m$ is estimated using the empirical mean. Then, we take the first absolute moment $A = \mathbb E[|X - \mathbb E[X]|$ and the standard deviation $\sqrt{\mathrm{Var}(X)}$. We find their expression using parameters. Their ratio depends only on the shape $a$, not on scale $b$ or location $m$. 
%This method has smaller limiting covariance matrix than the original \textsc{MME} \gr{for values of the shape $a$ and scale $b$ parameters that arise in financial modelling, the most common application of the GAL distribution}. 
Through our simulation study, we demonstrate that this method improves on the original \textsc{MME} (in terms of lower bias and mean square error) across a broad range of parameter values of $a$ and $b$. Of particular note is that performance is significantly improved for parameter values encountered in financial modelling, the most common application of the variance-gamma distribution. 
%However, we stress that caution should also be applying this estimator, because, like the classic MME, performance degrades to an unacceptably low standard as the shape parameter $a$ increases beyond 3.

% for values of the shape $a$ and scale $b$ parameters that arise in financial modelling, the most common application of the GAL distribution.

We also compare this modified MME to the classic MLE. As expected, for most parameter constellations MLE outperforms our modified MME (in terms of smaller bias and mean squared error), although  our modified MME is still quite competitive, and much more so than classic MME. %Interestingly, in our finite sample size simulations the modified MME performs better (in terms of bias and mean squared error) under some parameter constellations; in particular, for small values of the parameter $a$, although the MLE performs better for larger $a$. 
We also remark that \textsc{MLE} is difficult to implement in practice; see \cite{ct17} for an investigation of the computational problems associated with implementing MLE for parameter estimation for the variance-gamma distribution. 
%We also remark that the simulation results for MME reported in this paper were carried out using \emph{Mathematica}. We found that implementing MLE in \emph{Mathematica} was very slow and performance was unacceptable, with the algorithm often failing to even converge. We used the methods \emph{FindMaximum} and \emph{NMaximize} with various parameter configurations. Our computation of the MLE required numerical optimisation, performed with the Nelder-Mead algorithm as implemented in the R function \emph{optim}.

%In this paper, we take a well-known simple \textsc{MME}, which was assumed to work for \textsc{GAL} and \textsc{GNL} distributions. We use theory and simulation to show that it often does not work well even for symmetric versions of these distributions. Then we use absolute moments to improve efficiency for \textsc{GAL} across a wide range of parameter values that includes those encountered in financial modelling. Finally, we suggest some venues for further research. 

\subsection{Organization of this article} Section \ref{sec2} is devoted to the classic MME. We take the symmetric \textsc{VG} distribution, and state and prove asymptotic results and use them together with simulations to assess the performance of this MME. Section \ref{sec3} studies the modification of this MME with absolute moments for symmetric \textsc{VG} distributions. It is here that we make the main positive contribution of this note. 
%To simplify the theoretical analysis, in Section \ref{sec2}  we suppose that the location parameter $m = 0$ is known. However, in Section \ref{sec3} we no longer have this assumption. 
We perform the simulations both in case $m = 0$ and in the more realistic setting when $m$ is unknown. Section \ref{sec4} contains conclusions and some suggestions for subsequent research. Proofs are postponed until Appendix \ref{appendix}.

\section{Classic Method of Moments Estimation}\label{sec2}

\begin{figure}
\begin{center} 
\captionsetup{width=.9\textwidth}
 \includegraphics[width=11cm]{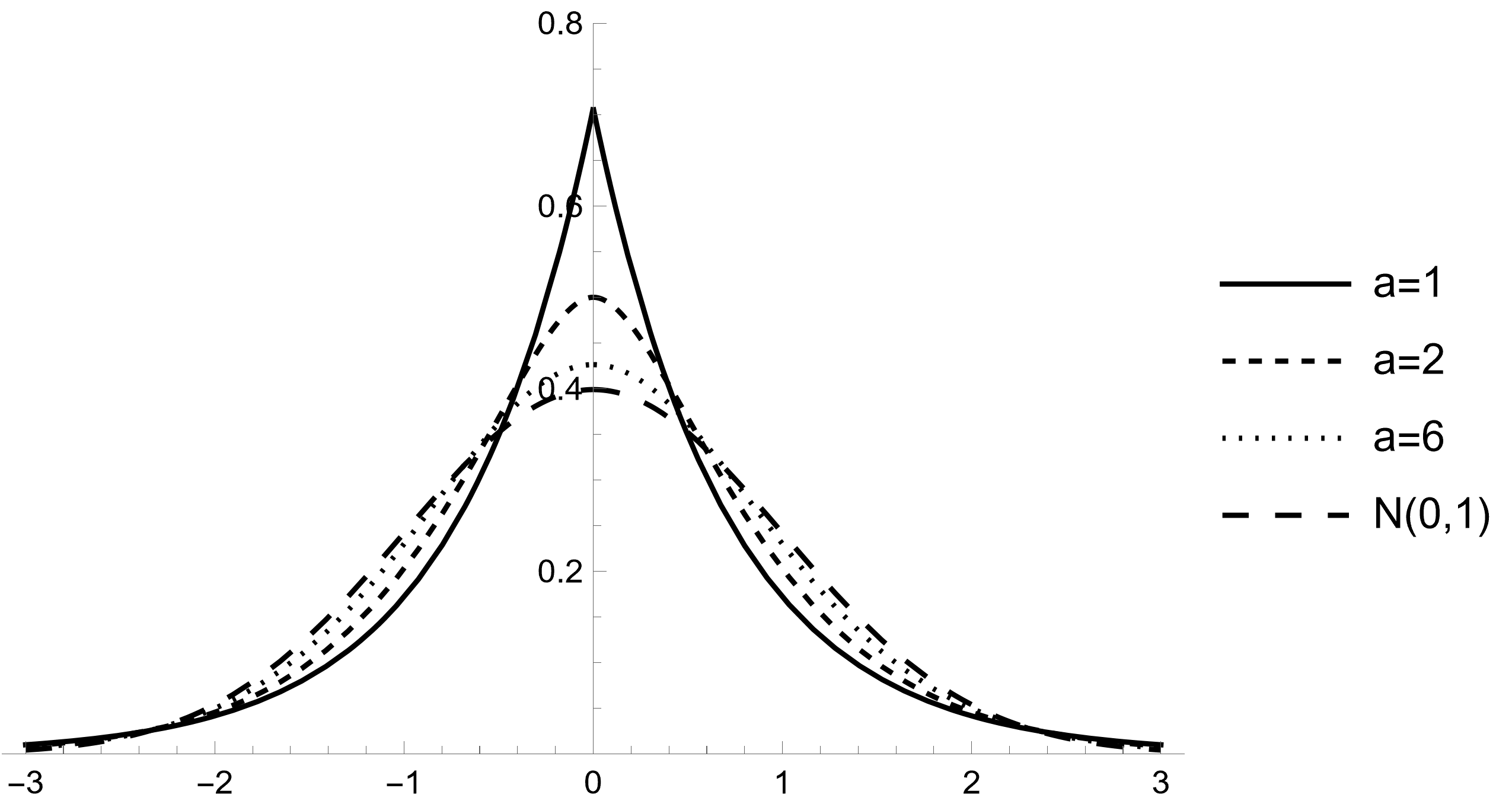}
\caption{The pdfs of the normal distribution and the generalized Laplace distribution for different values of $a$ and $b=1/a$.}
\label{fig:fig}
\end{center} 
\end{figure}

%\subsection{Symmetric Variance-Gamma} 
First, we define the family of distributions which we deal with. As discussed in the Introduction, we deal with a subset of the \textsc{GAL} family: {\it symmetric variance-gamma} (\textsc{SVG}) or {\it generalized Laplace}. We show the classic \textsc{MME} performs poorly for these distributions. The SVG distribution corresponds to setting $c=0$ in (\ref{eq:FT-VG}), and we have the following formula for the PDF: 
\begin{equation}\label{svgdefn}p(x) = \frac{1}{\sqrt{\pi b/2} \Gamma(a)}  \bigg(\frac{|x-m|}{\sqrt{2b}}\bigg)^{a-1/2} K_{a-1/2}\bigg(\frac{|x-m|}{\sqrt{b/2}} \bigg), \quad x\in\mathbb{R},
\end{equation}
where $K_\nu(x)=\int_0^\infty \mathrm{e}^{-x\cosh(t)}\cosh(\nu t)\,\mathrm{d}t$, $x>0$, is a modified Bessel function of the second kind \cite{olver}; note that this function is sometimes also referred to as the modified Bessel function of the third kind. A plot of the PDF for several parameter values is given in Figure \ref{fig:fig}. %, and the gamma function is defined by $\Gamma(y) := \int_0^{\infty}x^{y-1}e^{-x}\,\mathrm{d}x$, $y>0$. 
The parameterisation (\ref{svgdefn}) is very similar to those given by \cite{bibby,gaunt14,Book}, whilst the parametrisation of \cite{fs08} (similar to that of \cite{VG-Option}) is obtained via $a=\alpha$, $b=\sigma^2/a=\sigma^2/\alpha$ and $m=\mu$. When $a=1$, the PDF (\ref{svgdefn}) reduces to that of the classical Laplace distribution (see \cite{Book}).
%The parameter $a$ is a \emph{shape} parameter. As $a$ increases, the distribution becomes more rounded around its peak value $m$ (this can be seen from (\ref{pmutend}) below). The parameter $b$ is a \emph{scale} parameter. As $b$ decreases, the tails decay more quickly (see (\ref{pjv})).  The parameter $m$ is the \emph{location} parameter. The parameterisation (\ref{svgdefn}) is very similar to those given by \cite{bibby,gaunt14,Book}, whilst the parametrisation of \cite{fs08} (similar to that of \cite{VG-Option}) is obtained via $a=\alpha$, $b=\sigma^2/(2a)=\sigma^2/(2\alpha)$ and $m=\mu$. When $a=1$, the PDF (\ref{svgdefn}) reduces to that of the classical Laplace distribution (see \cite{Book}).
The parameter $m$ is the \emph{location} parameter. Since the distribution is symmetric around $m$, this is the mean and the median. In addition, this is the (only) mode: The symmetric variance-gamma distribution is unimodal. The parameter $a$ is the \emph{shape} parameter. As $a$ increases, the distribution becomes more {\it rounded} around its peak value $m$. For example, it is differentiable at $x = m$ for $a > 1$. For $a \in (1/2, 1]$, the density behaves as $\mathrm{const} \cdot |x - m|^{2a-1}$ as $x \to m$: continuous, but not differentiable. Next, the density has a logarithmic singularity at $x=m$ if $a = 1/2$ and a power singularity if $a \in (0, 1/2)$. The parameter $b$ is the \emph{scale} parameter, and as $b$ decreases the tails decay more quickly. A recent article \cite{Gaunt} has precise statements and proofs of these asymptotic results.
%, as well as many other useful statements and an extensive bibliography. 

% \begin{figure}
% \begin{center} 
% \captionsetup{width=.9\textwidth}
%  \includegraphics[width=11cm]{pdf.png}
% \caption{The pdfs of the normal distribution $N(0,1)$ \includegraphics[scale=1]{red} and the generalized Laplace distribution for $a=1$  \includegraphics[scale=1]{blue}, $a=2$  \includegraphics[scale=1]{orange}, $a=6$ \includegraphics[scale=1]{green} and  $b=1/a$.}
% \label{fig:fig}
% \end{center} 
% \end{figure}

\begin{rmk}
\label{rmk:convergence-normal}
As $a \to \infty$, the symmetric variance-gamma distribution converges weakly to the normal distribution $\mathcal N(m, \sigma^2)$ if we scale the parameter $b$ as $b = \sigma^2/a$. This easily follows from the convergence of characteristic function given in (\ref{eq:FT-VG}) to the characteristic function $\exp(im\omega - \sigma^2\omega^2/2)$ of the normal distribution $\mathcal N(m, \sigma^2)$. 
\end{rmk}

The SVG distribution has a fundamental representation in terms of gamma and normal random variables (see \cite[Proposition 4.1.2]{Book}). 
Fix parameters $a, b > 0$. Then a SVG random variable $X$ can be written as
\begin{equation}\label{vgrep}X =_d \sqrt{bG}Z,
\end{equation}
where $G \sim \Gamma(a, 1)$ and $Z \sim \mathcal N(0, 1)$ are independent.

Of course, all centered odd moments are zero: $\mathbb E[(X - \mathbb E X)^{2n+1}] = 0$ for $n = 0, 1, 2, \ldots$. With the representation (\ref{vgrep}), moments of the SVG can be calculated using standard formulas for the moments of the gamma and normal distributions (see \cite[Proposition 4.1.6]{Book}). In particular, the variance and fourth central moment are \begin{equation}
\label{eq:moments}
V = \mathrm{Var}(X) = ab,\quad K = \mathbb E[(X - \mathbb E X)^4] = 3a(a+1)b^2.
\end{equation}
With the formulas in (\ref{eq:moments}), we can develop a classic method of moments. We shall show it is consistent and asymptotically normal, but with large asymptotic variance. Assume $X_1, X_2, \ldots$ are i.i.d. from this variance-gamma distribution~\eqref{vgrep}. Define the empirical second and fourth moments: 
\begin{equation}
\label{eq:empirical-moments}
\hat{V} := \frac1N\sum\limits_{i=1}^N(X_i - \overline{X})^2\quad \mbox{and}\quad \hat{K} := \frac1N\sum\limits_{i=1}^N(X_i - \overline{X})^4. 
\end{equation}

\begin{thm}\label{thm1} The MME estimators are given by 
\begin{equation}
\label{eq:mme}
\hat{m} := \overline{X},\quad \hat{a} := \frac{3\hat{V}^2}{\hat{K} - 3\hat{V}^2},\quad \hat{b} := \frac{\hat{K}}{3\hat{V}} - \hat{V}.
\end{equation}
They are consistent: $(\hat{a}, \hat{b}, \hat{m}) \to (a, b, m)$ almost surely as $N \to \infty$.
\end{thm}

\begin{rmk} In (\ref{eq:empirical-moments}), we have used the naive moment estimators rather than the unbiased ones; for example, multiplication of $\hat{V}$ by a factor of $N/(N-1)$ yields an unbiased estimator \gr{of variance}. We note, however, that our use of the naive estimators will have only a negligible effect for the purposes of this article, as, by the Slutsky theorem, this slight changes preserves consistency and asymptotic normality, and does not change limiting covariance matrix of the estimators. Also, for our simulation studies, we take a sample size of $N=1000$, for which there is little difference in performance between the naive and unbiased estimators.  
%For Theorem~\ref{thm:final}, in place of $\hat{V}'$, we can use $s^2$, the classic empirical variance. The only difference from $\hat{V}'$ is division by $n-1$ instead of $n$. By the Slutsky theorem, this slight change preserves consistency and asymptotic normality, and does not change limiting covariance matrix.
\end{rmk}

\begin{rmk}The moment estimators of Theorem \ref{thm1} do not satisfy support constraints on the parameters, because, for given data, we do not necessarily have $\hat{K}>3\hat{V}^2$. This condition is of course met (with high probability; see Lemma \ref{lemkv} below) for large sample sizes, because the estimators are consistent and $K-3V^2=3ab^2>0$. The fact that method of moments estimators do not always satisfy support constraints is a well-known deficiency of the method; we also remark that our simulation results suggest that classical MME is not suitable for small sample sizes (when there is a possibility of the support conditions being violated). %However, it is easy to show that  $\mathbb P(\hat{K} \le 3\hat{V}^2) \to 0$ as $N \to \infty$; so for large sample size, we can apply MME with probability close to 1. Similar results apply to the modified MME in the next section. 
\end{rmk}

\begin{lemma}\label{lemkv}
There exists a constant $C > 0$, dependent upon $a$ and $b$, such that 
$$
\mathbb P(\hat{K} > 3\hat{V}^2) \ge 1 - \frac CN.
$$
\end{lemma}

The theoretical result of Lemma \ref{lemkv} is complemented with a simulation study that gives estimates on the probability of the support condition $\hat{K} > 3\hat{V}^2$ being met (in the case of unknown $m$); see Table \ref{tablesupt} below.

\begin{rmk}
For the normal distribution, we have $K = 3\sigma^4 = 3V^2$. In light of the Remark~\ref{rmk:convergence-normal}, we can expect that with high probability, we cannot apply the MME for large $a$. We will see later in simulations that, in general, the MME works better for smaller $a$. 
\label{rmk:failure-normal}
\end{rmk}

As we shall see, the limiting variance of each estimator is rather large. 
To illustrate this, let us consider a special case, where, without loss of generality, we have $m = 0$. 
Then we can modify the estimators~\eqref{eq:empirical-moments} for $V$ and $K$:
\begin{equation}
\label{eq:modifications}
\hat{V}' %= s^2
 = \frac1N\sum\limits_{j=1}^NX_j^2\quad \mbox{and}\quad \hat{K}' = \frac1N\sum\limits_{j=1}^NX_j^4.
\end{equation}
\begin{rmk} Below, we always add primes to empirical moments if we do NOT subtract the empirical mean $\overline{X}$. 
However, for empirical centered moments, where we do subtract $\overline{X}$, we do not add primes. 
\end{rmk}
We plug~\eqref{eq:modifications} into formulas~\eqref{eq:mme} for 
$\hat{a}$ and $\hat{b}$ and get consistent estimators for $a$ and $b$. Below we state an asymptotic normality result. \gr{We note that in our result we are able to give an explicit formula for the limiting covariance matrix; we should, however, point out that the asymptotic covariance matrix of the MLE is complicated and, to our best knowledge, an explicit formula is not available in the literature.}
%; the proof is given in Appendix \ref{appendix}.
% We might as well have done this for general unknown $m$ and original estimators~\eqref{eq:mme} using~\eqref{eq:empirical-moments}. But this would complicate the analysis. It is quite enough to consider this simple case. Even here, the limiting variance is unacceptably large. 

\begin{thm}\label{thm:original}
Assuming $m = 0$, the modified estimators from~\eqref{eq:modifications} are asymptotically normal: 
\begin{align*}
\begin{split}
&\sqrt{N}[(\hat{a}, \hat{b}) - (a, b)]  \to_d \mathcal N_2\left([0,  0], \Sigma\right) \quad \mbox{with}\\
&\Sigma=\frac{1}{3}\begin{bmatrix}
2(4a^4+36a^3+95a^2+63a) & -2(4a^3+36a^2+101a+69)b^2 \\ -2(4a^3+36a^2+101a+69)b^2 & (8a^2+72a+220+159a^{-1})b^4 
\end{bmatrix}.
\end{split}
\end{align*}
%\begin{align*}
%\begin{split}
%\sqrt{N}[(\hat{a}, \hat{b}) - (a, b)]  &\to \mathcal N_2\left(
%\begin{bmatrix}0 \\ 0 \end{bmatrix}, \Sigma\right)\quad \mbox{with}\quad \Sigma = JCJ^T;\\
%J = \begin{bmatrix}
%\frac{2(a+1)}{b^{2}} & -\frac1{3b^4}\\ -2-\frac1a & \frac1{3ab^2} 
%\end{bmatrix},
%\quad 
%C &= \begin{bmatrix}
%(2a^2 + 3a)b^4 & a(a+1)(12a + 30)b^6 \\ a(a+1)(12a+30)b^6 & a(a+1)(96a^2 + 516a + 630)b^8 
%\end{bmatrix}.
%\end{split}
%\end{align*}
\end{thm}

\begin{figure}
\centering
\subfloat{\includegraphics[width=0.45\textwidth]{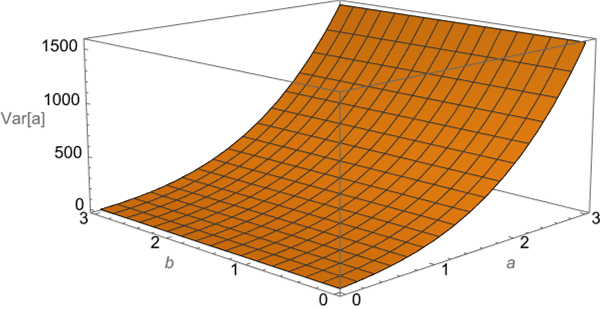}}
\subfloat{\includegraphics[width=0.45\textwidth]{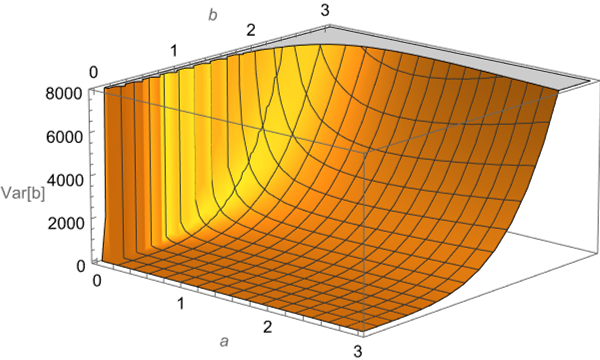}}
\caption{Entries $(\Sigma)_{a,a}$ and $(\Sigma)_{b,b}$ in the asymptotic covariance matrix $\Sigma$.}
\label{fig:variances}
\end{figure}

\begin{figure}\label{fig2}
  \includegraphics[width=0.5\linewidth]{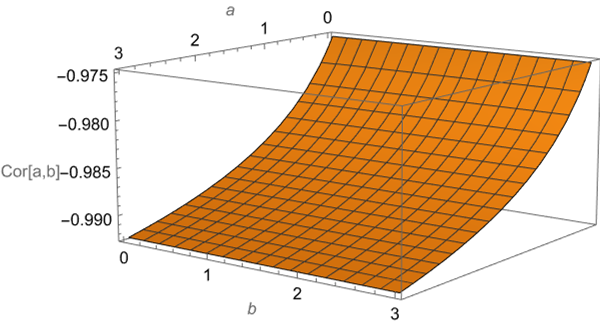}
  \caption{Asymptotic correlation.}
  \label{fig3}
\end{figure}

%Unfortunately, the limiting covariance matrix $\Sigma$ is rather large. 
A plot of the entries in the covariance matrix $\Sigma$ are given in Figures \ref{fig:variances} and \ref{fig3}. It can be seen that the variance of the estimator 
$\hat{a}$ get large as $a$ increases, whilst the variance of the estimator $\hat{b}$ becomes large as $b$ increases. 
%This is because $\mathbb E[X^6]$ and especially $\mathbb E[X^8]$ are large, particularly for large values of $a$ and $b$. 
This means estimators $\hat{a}$ and $\hat{b}$ may often be of poor quality, particularly for moderate sample sizes or for larger values of $a$ and $b$. 

\begin{exm} Take $a = b = 1$. Then 
\[\Sigma = \begin{bmatrix} 132 & -140 \\ -140 & 153 \end{bmatrix}.\]
%$$
%J = 
%\begin{bmatrix}
%4 & -\frac13 \\ -3 & \frac13
%\end{bmatrix}
%\quad \mbox{and}\quad 
%C = 
%\begin{bmatrix}
%5 & 84 \\ 84 & 2484
%\end{bmatrix}
%\quad \mbox{therefore}, \quad \Sigma = JCJ^T = \begin{bmatrix} 132 & -140 \\ -140 & 153 \end{bmatrix}.
%$$
For example, if $N = 100$, then the standard deviation of the estimator $\hat{b}$ is $\sqrt{153}/10 > 1$. This implies that the estimator is of low quality. Note also that the limiting correlation is $-140/\sqrt{132\cdot 153} = -0.985$, which is very strong. 
\label{exm:classic}
\end{exm}

We now state an asymptotic normality result for the more realistic case of unknown $m$.

\begin{thm}\label{thm:final0} The estimators $(\overline{X}, \hat{a}, \hat{b})$, defined in (\ref{eq:mme}), are asymptotically normal. The limiting covariance matrix is 
$\mathrm{diag}(ab, \Sigma)$, where $\Sigma$ is the limiting covariance matrix from Theorem~\ref{thm:original}. 
\end{thm}

\begin{rmk}The limiting covariance $3\times 3$ matrix is block diagonal. Therefore, asymptotically, the estimators for $(\hat{a}, \hat{b})$ and the empirical mean are independent.
\end{rmk}

In addition, we performed simulations to assess the quality of the MME~\eqref{eq:mme}: Fix $a, b > 0$ and repeat $k=10,000$ times the following procedure. We implemented the procedure in \emph{Mathematica}.

\begin{enumerate}
\item Repeat $k=10,000$ times the following procedure: 
\begin{itemize}
\item Generate a symmetric variance-gamma sample of $n$ variables with $m = 0$.
\item Compute estimates from~\eqref{eq:mme}.
\item Denote them to be $(\hat{a}_i, \hat{b}_i)$ for the $i$th iteration. 
\end{itemize}
\item Find the average over all estimates $\mathrm{Mean}_a := \frac1k\sum_{i=1}^k\hat{a}_i$ and $\mathrm{Mean}_b := \frac1k\sum_{i=1}^k\hat{b}_i$. 
\item Find mean squared errors $\mathrm{MSE}_a = \frac1{k}\sum_{i=1}^k(\hat{a}_i - a)^2$ and $\mathrm{MSE}_b = \frac1{k}\sum_{i=1}^k(\hat{b}_i - b)^2$. 
\end{enumerate} 

\begin{center}
\begin{table}\scriptsize
\centering
\begin{tabular}{C|C|C|C|C|C|C|C}
a & b & \text{Bias}(\hat{a}) & \text{MSE}(\hat{a}) & \text{StError}(\hat{a})  & \text{Bias}(\hat{b}) & \text{MSE}(\hat{b}) & \text{StError}(\hat{b})\\ 
\hline
 0.25 & 0.01 & 4.45\text{e-2} & 9.64\text{e-3} & 2.77\text{e-3} & -3.34\text{e-4} & 2.57\text{e-5} & 1.6\text{e-4} \\
 0.25 & 0.1 & 4.4\text{e-2} & 9.65\text{e-3} & 2.78\text{e-3} & -3.41\text{e-3} & 2.51\text{e-3} & 1.58\text{e-3} \\
 0.25 & 1 & 4.41\text{e-2} & 9.65\text{e-3} & 2.78\text{e-3} & -3.94\text{e-2} & 2.2\text{e-1} & 1.48\text{e-2} \\
 0.25 & 5 & 4.33\text{e-2} & 9.48\text{e-3} & 2.76\text{e-3} & -1.78\text{e-1} & 5.67\text{e}\text{} & 7.51\text{e-2} \\
 \hline
 0.5 & 0.01 & 7.06\text{e-2} & 3.11\text{e-2} & 5.11\text{e-3} & -2.7\text{e-4} & 1.63\text{e-5} & 1.28\text{e-4} \\
 0.5 & 0.1 & 6.81\text{e-2} & 3.07\text{e-2} & 5.11\text{e-3} & -2.61\text{e-3} & 1.54\text{e-3} & 1.24\text{e-3} \\
 0.5 & 1 & 7.05\text{e-2} & 3.1\text{e-2} & 5.1\text{e-3} & -2.8\text{e-2} & 1.64\text{e-1} & 1.28\text{e-2} \\
 0.5 & 5 & 6.82\text{e-2} & 3.1\text{e-2} & 5.13\text{e-3} & -1.14\text{e-1} & 4.08\text{e}\text{} & 6.38\text{e-2} \\
 \hline
 1 & 0.01 & 1.31\text{e-1} & 1.29\text{e-1} & 1.06\text{e-2} & -2.13\text{e-4} & 1.4\text{e-5} & 1.18\text{e-4} \\
 1 & 0.1 & 1.33\text{e-1} & 1.3\text{e-1} & 1.06\text{e-2} & -2.41\text{e-3} & 1.34\text{e-3} & 1.15\text{e-3} \\
 1 & 1 & 1.24\text{e-1} & 1.25\text{e-1} & 1.05\text{e-2} & -1.29\text{e-2} & 1.47\text{e-1} & 1.21\text{e-2} \\
 1 & 5 & 1.24\text{e-1} & 1.27\text{e-1} & 1.06\text{e-2} & -6.47\text{e-2} & 3.59\text{e}\text{} & 5.99\text{e-2} \\
 \hline
 2 & 0.01 & 2.98\text{e-1} & 7.84\text{e-1} & 2.64\text{e-2} & -1.28\text{e-4} & 1.46\text{e-5} & 1.21\text{e-4} \\
 2 & 0.1 & 3.08\text{e-1} & 8.01\text{e-1} & 2.66\text{e-2} & -1.36\text{e-3} & 1.54\text{e-3} & 1.24\text{e-3} \\
 2 & 1 & 3.22\text{e-1} & 8.07\text{e-1} & 2.65\text{e-2} & -2.12\text{e-2} & 1.48\text{e-1} & 1.21\text{e-2} \\
 2 & 5 & 3.09\text{e-1} & 8.19\text{e-1} & 2.69\text{e-2} & -5.48\text{e-2} & 4.06\text{e}\text{} & 6.37\text{e-2} \\
 \hline
 3 & 0.01 & 6.52\text{e-1} & 4.78\text{e}\text{} & 6.6\text{e-2} & -2.15\text{e-4} & 1.75\text{e-5} & 1.32\text{e-4} \\
 3 & 0.1 & 7.34\text{e-1} & 6.63\text{e}\text{} & 7.8\text{e-2} & -2.78\text{e-3} & 1.75\text{e-3} & 1.32\text{e-3} \\
 3 & 1 & 6.7\text{e-1} & 4.12\text{e}\text{} & 6.06\text{e-2} & -1.72\text{e-2} & 1.88\text{e-1} & 1.37\text{e-2} \\
 3 & 5 & 6.83\text{e-1} & 8.46\text{e}\text{} & 8.94\text{e-2} & -1.06\text{e-1} & 4.49\text{e}\text{} & 6.69\text{e-2}
 \end{tabular}
\caption{\label{table_mom} Simulation results for classic MME with known $m=0$ and sample size $N=1000$.}
\end{table}
\end{center}

\begin{center}
\begin{table}\scriptsize
\centering
\begin{tabular}{C|C|C|C|C|C|C|C}
a & b & \text{Bias}(\hat{a}) & \text{MSE}(\hat{a}) & \text{StError}(\hat{a})  & \text{Bias}(\hat{b}) & \text{MSE}(\hat{b}) & \text{StError}(\hat{b})\\ 
 \hline
 0.25 & 0.01 & 1.31\text{e-1} & 3.25\text{e-2} & 3.92\text{e-3} & -3.48\text{e-3} & 2.39\text{e-5} & 1.09\text{e-4}  \\ 
 0.25 & 0.1 & 8.33\text{e-2} & 2.06\text{e-2} & 3.69\text{e-3} & -2.21\text{e-2} & 1.55\text{e-3} & 1.03\text{e-3} \\ 
 0.25 & 1 & 5.31\text{e-2} & 1.3\text{e-2} & 3.19\text{e-3} & -1.4\text{e-1} & 1.03\text{e-1} & 9.13\text{e-3} \\ 
 0.25 & 5 & 3.74\text{e-2} & 9.14\text{e-3} & 2.78\text{e-3} & -4.92\text{e-1} & 1.94\text{e} & 4.12\text{e-2}  \\
  \hline
 0.5 & 0.01 & 2.19\text{e-3} & 1.25\text{e-3} & 1.12\text{e-3} & 8.2\text{e-4} & 8.59\text{e-4} & 9.27\text{e-4}  \\ 
 0.5 & 0.1 & 2.73\text{e-3} & 1.01\text{e-3} & 10\text{e-4} & 1.62\text{e-4} & 1.41\text{e-4} & 3.76\text{e-4}  \\ 
 0.5 & 1 & 3.07\text{e-3} & 1.03\text{e-3} & 1.01\text{e-3} & 8.43\text{e-4} & 1.44\text{e-2} & 3.79\text{e-3}  \\ 
 0.5 & 5 & 3.38\text{e-3} & 1.04\text{e-3} & 1.01\text{e-3} & -5.43\text{e-3} & 3.56\text{e-1} & 1.89\text{e-2}  \\ 
  \hline
 1 & 0.01 & 1.97\text{e-2} & 1.51\text{e-2} & 3.84\text{e-3} & -3.63\text{e-5} & 2.35\text{e-6} & 4.85\text{e-5}  \\ 
 1 & 0.1 & 1.74\text{e-2} & 1.45\text{e-2} & 3.77\text{e-3} & -9.45\text{e-5} & 2.33\text{e-4} & 4.83\text{e-4}  \\ 
 1 & 1 & 1.64\text{e-2} & 1.42\text{e-2} & 3.73\text{e-3} & 2.9\text{e-4} & 2.29\text{e-2} & 4.79\text{e-3}  \\ 
 1 & 5 & 1.79\text{e-2} & 1.48\text{e-2} & 3.8\text{e-3} & -1.19\text{e-2} & 5.83\text{e-1} & 2.41\text{e-2}  \\ 
  \hline
 2 & 0.01 & 1.03\text{e-1} & 2.35\text{e-1} & 1.5\text{e-2} & 4.63\text{e-5} & 1.13\text{e-5} & 1.06\text{e-4}  \\ 
 2 & 0.1 & 9.94\text{e-2} & 2.29\text{e-1} & 1.48\text{e-2} & -1.3\text{e-4} & 5.01\text{e-4} & 7.08\text{e-4}  \\ 
 2 & 1 & 9.54\text{e-2} & 2.35\text{e-1} & 1.5\text{e-2} & 3\text{e-4} & 4.93\text{e-2} & 7.02\text{e-3}  \\ 
 2 & 5 & 9.3\text{e-2} & 2.27\text{e-1} & 1.48\text{e-2} & 1.01\text{e-2} & 1.26\text{e} & 3.55\text{e-2}  \\ 
  \hline
 3 & 0.01 & 2.8\text{e-1} & 1.21\text{e} & 3.36\text{e-2} & -1.59\text{e-5} & 8.19\text{e-6} & 9.05\text{e-5}  \\ 
 3 & 0.1 & 2.73\text{e-1} & 1.23\text{e} & 3.39\text{e-2} & 4.33\text{e-5} & 8.29\text{e-4} & 9.1\text{e-4} \\
 3 & 1 & 2.89\text{e-1} & 1.26\text{e} & 3.42\text{e-2} & -4.46\text{e-3} & 8.19\text{e-2} & 9.05\text{e-3} \\ 
 3 & 5 & 2.87\text{e-1} & 1.23\text{e} & 3.39\text{e-2} & -1.87\text{e-2} & 2.05\text{e} & 4.53\text{e-2} 
 \end{tabular} 
\caption{\label{table_mle} Simulation results for classic MLE with known $m=0$ and sample size $N=1000$.}
\end{table}
\end{center}

%We performed this for $k = 500$, $n = 1000$, choosing $a = 3$ and $b = 1$. Results: the mean of $a$ is $3.5765$, the mean of $b$ is $2.9569$;   the MSE for $a$ is $2.8928$, the MSE for $b$ is $4.3255$.  This is another illustration that the method of moment estimation tends to fall apart and estimates from the method deviate wildly from the original parameters.

\smallskip
The results are reported in Table \ref{table_mom}. We chose a wide range of parameter values, under which the SVG distribution exhibits quite different behaviour. Indeed, as mentioned above, $a=0.25$ ($a=0.5$) corresponds to a power law (logarithmic) singularity in the PDF at $x=m=0$, whilst $a=1$ corresponds to the classical Laplace distribution, and for $a\geq2$ the distribution becomes increasingly rounded about the peak. Similarly, we consider various different orders of magnitude for the scale parameter $b$.   Later, we repeat the simulations in the more realistic setting, where the true value of the location parameter $m=0$ is unknown. 

As a means of comparison, we repeated the simulations using classic MLE. In order to calculate the MLE, we used the Nelder-Mead algorithm, as implmented in the R function \emph{optim}. For the evaluation of the log-likelihood function, we used the  \emph{dvg} function delivered by the R \emph{VarianceGamma} package. In order to avoid the singularity of the density and negative values for the parameters, we set $a$ or $b$ equal to $10^{-5}$ when the Nelder-Mead algorithm wants to evaluate the log-likelihood function at negative parameter values, and similarly we set the variables $\sigma=\sqrt{ab}$ and $\nu=1/a$ equal to $10^{-4}$ as soon as they are smaller than $10^{-4}$. The results are reported in Table \ref{table_mle}.

From Table \ref{table_mom}, we infer that for small values of $a$ and $b$ the performance of the estimators in terms of bias and MSE is quite reasonable. In fact, in our finite sample simulation, when $a=0.25$, the MSE of the moment estimator $\hat{a}$ is smaller than the MSE for the corresponding MLE estimator for all values of $b$ except $b=5$. This good performance for small $a$, is consistent with results in the literature in which MME has been used to fit the variance-gamma distribution to financial data, like log returns of assets prices over many ($1000+$) trading days; see, for example \cite{VG} in which values of the order $b=2\times 10^{-5}$ and $a\leq0.5$ were considered. Indeed, when fitting the VG distribution to log returns of financial assets, very small values of $b$ and values of $a\leq2$ are typically encountered. However, as the values of $a$ and $b$ increase, we see that the performance of the classic MME degrades. 
%At a sample size of $N=1000$, we see that, for all values of $a$ considered, the MSE of $\hat{b}$ is unacceptably large for unknown $b=5$.  Moreover, when $a\geq1$ we see that the MSE of $\hat{b}$ is often unacceptably large across the whole range of values of $b$ considered. 
We see that the MSE of the moment estimator $\hat{a}$ when $a=3$ is now significantly larger than for the MLE. We found that this gets worse as $a$ increases beyond $3$. We therefore conclude, that whilst performance of the MME in terms of bias and MSE is quite reasonable for some of the specialised values of $a$ and $b$ that have been tested on some financial data sets, for other values of $a$ and $b$ performance deteriorates, meaning that the performance of the method is quite sensitive to the true values of model parameters. We do, however, note that as $a$ becomes larger, the SVG PDF approaches that of the normal distribution (see Remark \ref{rmk:convergence-normal} and Figure \ref{fig:fig}), and so for practical purposes the differences in performance between classic MME and MLE for $a\geq3$ may not be as great as suggested by our simulations. 

%As such, researchers should exercise caution when implementing classic MME.

\section{Modified Method of Moments}\label{sec3}

\subsection{Main idea} 
Define the {\it absolute centered moment} $A := \mathbb E|X - \mathbb E[X]|$.  The following formula for $A$ was stated and proved in \cite[Proposition 2.2]{Gaunt}:
\begin{equation}
%\label{eq:first-moment}
\label{eq:absolute-moments}
A = \mathbb E|X - \mathbb E[X]| = \left(\frac{2b}{\pi}\right)^{0.5}\frac{\Gamma(a+0.5)}{\Gamma(a)}.
\end{equation}
Knowing $A$ and $V$, we can solve for $a$:
$$
\frac{\sqrt{V}}{A} = \sqrt{\frac{\pi}2}
\frac{\sqrt{a}\cdot\Gamma(a)}{\Gamma(a+0.5)}.
$$
Taking logarithms, we get: 
$$
0.5\ln V - \ln A = L(a) := 0.5\ln(\pi/2) + 0.5\ln a + \ln\Gamma(a) - \ln\Gamma(a+0.5).
$$

\begin{lemma}
The function $L$ is a one-to-one strictly decreasing smooth mapping from $(0, \infty)$ onto $(0.5\ln(\pi/2), \infty)$, with $L'(x) < 0$ for $x > 0$. 
\label{lemma:decreasing}
\end{lemma}

\subsection{Centered case} First, assume that we know the mean (and median) is zero. Then we no longer have to subtract the empirical mean from each observation. This makes proofs of consistency and asymptotic normality for this case easier, and they serve as stepping stone for proofs in the general  case. 

%Centrality seems like an artificial restriction. However, this special case is also important in practice. For example, in the third author's financial econometrics research, we fitted linear regressions with heavy-tailed residuals. For said residuals, we successfully used symmetric variance-gamma distributions. However, by construction residuals had zero mean, since a regression has an intercept, which is an additive constant.

Centrality seems like an artificial restriction. However, this special case also arises in practice. For example, consider linear regressions with residuals with heavier than Gaussian tails. For said residuals, one could fit symmetric variance-gamma distributions. However, by construction residuals have zero mean, since a regression has an intercept, which is an additive constant.

In the next subsection, we consider the general case. Meanwhile, here we used simplified estimators for the first absolute moment and the second moment (which is equal to the variance). 

We can estimate $A$ from~\eqref{eq:absolute-moments} as
\begin{equation}
\label{eq:empirical-A}
\hat{A}' := \frac1N\sum\limits_{j=1}^N|X_j|.
\end{equation}
The estmate $\hat{V}'$ of $V$ is given in~\eqref{eq:modifications}. Lemma~\ref{lemma:decreasing} allows us to estimate $a$ as follows: We can define the inverse function 
$$
\ell : (0.5\ln(\pi/2), \infty) \to (0, \infty)
$$ 
to $L$. It is continuous and smooth. Thus we can define the estimators for $a$ and $b$:
\begin{equation}
\label{eq:new-est}
\gr{\hat{a}'} := \ell(0.5\ln\hat{V}' - \hat{A}'),\quad \hat{b} := \frac{\hat{V}'}{\hat{a}'}.
\end{equation}
%The following theorem is proved in Appendix \ref{appendix}.

\begin{thm}\label{normod}
The estimators in~\eqref{eq:new-est} are consistent and asymptotically normal. 
\end{thm}

\begin{rmk}
    \gr{The covariance matrix for the asymptotic normal distribution in Theorem \ref{normod} is not explicit, and so is not reported.}
\end{rmk}

As for any MME, the feasibility question arises. We have a result similar to Lemma \ref{lemkv} above.

\begin{lemma}\label{lemma:applicability-estimate}
There exists a constant $C > 0$, dependent upon $a$ and $b$, such that the modified MME can be, in fact, applied with probability at least $1 - C/N$:
$$
\mathbb P\left(0.5\ln\hat{V}' - \hat{A}' > 0.5\ln(\pi/2)\right) \ge 1 - \frac CN.
$$
\end{lemma}

\begin{rmk}  The MLE for the (centered) symmetric Laplace distribution $f(x) = 0.5\lambda \exp(-\lambda|x|)$, $x\in\mathbb{R}$, is just our modified MME: It uses only $A = \mathbb E|X| = \lambda^{-1}$, since this is a one-parameter family of distributions. Classic MME uses only classic moments $\mathbb E[X^{2k}]$. Modified MME uses $\mathbb E[|X|^n]$. The classic MME for this symmetric Laplace family will use 
$\mathbb E[X^2] = 2\lambda^{-2}$. The resulting estimate is consistent and asymptotically normal, but not efficient. This observation is a motivation for using the modified MME for generalized symmetric Laplace (symmetric variance-gamma) distributions.
\end{rmk}

\begin{exm}
We now compute the asymptotic variance for the estimators~\eqref{eq:new-est} in the case $a = b = 1$, to compare it with the classic MME. First, in this case $A = 1/\sqrt{2}$ and $V = 1$, 
therefore $L(1) = 0.5\ln V - \ln A = \ln\sqrt{2}$ and $\ell(\ln\sqrt{2}) = 1$. Using Python, we compute $-L'(1) = \alpha = 0.1137$ and therefore $\ell'(\ln\sqrt{2}) = -\alpha^{-1}$. 
 Next, the gradient of the function $(x, y) \mapsto 0.5\ln y - \ln x$ (the first component of $\Phi$) is equal to $(-1/x, 0.5/y)$. At $x = A$ and $y = V$ this is equal to $z = (-\sqrt{2}, 0.5)$. 
Thus the gradient of the function $\ell(0.5\ln y - \ln x)$ is equal to $w = -\alpha^{-1}z = (\sqrt{2}\alpha^{-1}, -0.5\alpha^{-1})$. 
The Jacobi matrix $J_2$ of the function $\Psi$ at $x = 1/\sqrt{2}$ and $y = 1$ has first row equal to $w$. The second row can be computed similarly:
$$
J_2 = \begin{bmatrix}
\sqrt{2}\alpha^{-1} & -0.5\alpha^{-1}\\
-\sqrt{2}\alpha^{-1} & 1 + 0.5\alpha^{-1} 
\end{bmatrix}.
$$
From~\eqref{eq:classic}, we have 
$$
\sqrt{N}\left[(\hat{A}', \hat{V}') - (A, V)\right] \rightarrow_d \mathcal N_2([0, 0], C),\quad C_2 = \begin{bmatrix} 1 & 3/\sqrt{2}\\ 3/\sqrt{2} & 6 \end{bmatrix}.
$$
By the delta method, the limiting covariance matrix is 
$$
\Sigma_2 = J_2C_2J^T_2 = 
\begin{bmatrix}
38.67 & -38.67 \\
-38.67 & 44.67 
\end{bmatrix}.
$$
Both limiting variances are smaller than those in Example~\ref{exm:classic}. To be fair, the limiting correlation is still $\approx -93\%$, which is very strong. This is perhaps to be expected since $\hat{b} = \hat{V}/\hat{a}$. 
\end{exm}

\subsection{General case} Now, we drop the assumption that $m = 0$.
%: symmetric variance-gamma distribution is centered. 
We have the obvious estimator $\hat{m} = \overline{X}$ for the location parameter $m$. We subtract this from each observation $X_i$ to modify the estimator $\hat{A}'$ from~\eqref{eq:empirical-A} of the absolute first moment:
\begin{equation}
\label{eq:modified-A}
\hat{A} := \frac1N\sum\limits_{k=1}^N|X_k - \overline{X}|.
\end{equation}
For empirical variance,  we use $\hat{V}$ from~\eqref{eq:empirical-moments}:
$$
\hat{V} := \frac1N\sum\limits_{k=1}^N(X_k - \overline{X})^2. 
$$
We create estimators $\hat{a}, \hat{b}$ from $\hat{A}, \hat{V}$ just like in~\eqref{eq:new-est}. 

\begin{thm}\label{thm:final} The estimators $(\overline{X}, \hat{a}, \hat{b})$ are consistent and asymptotically normal. The limiting covariance matrix is 
$\mathrm{diag}(ab, \Sigma)$, where $\Sigma$ is the limiting covariance matrix from Theorem~\ref{normod}. 
\end{thm}

We have the following analogue of Lemma \ref{lemma:applicability-estimate} for the case \gr{where $m$ is unknown}. Simulations were used to estimate the probability that $0.5\ln\hat{V} - \ln\hat{A} > 0.5\ln(\pi/2)$, with the results reported in Table \ref{tablesupt}. We see similar behaviour for the probabilities for both the classic and modified MME estimates, although the probability of the classic moment estimators existing is greater than for the modified MME. Note that these simulations have been performed on small sample sizes; for a sample size of $N=1000$ (for the parameter values we considered) the estimators exist with very high probability.

\begin{lemma}\label{lemma:feasibility-estimate}
There exists a constant $C > 0$ (possibly different from the one in Lemma~\ref{lemma:applicability-estimate}), dependent upon $a$ and $b$, such that the modified MME can be, in fact, applied with probability at least $1 - C/N$:
$$
\mathbb P\left(0.5\ln\hat{V} - \ln\hat{A} > 0.5\ln(\pi/2)\right) \ge 1 - \frac CN.
$$
\end{lemma}

\begin{rmk}
Similarly to Remark~\ref{rmk:failure-normal}, we can note that for any normal distribution, $V = \sigma^2$ and $A = (2/\pi)^{1/2}\sigma$. Thus $0.5\ln V - \ln A = 0.5\ln(\pi/2)$. In light of Remark~\ref{rmk:convergence-normal}, we can expect that the modified MME (similarly to the classic MME) is impossible to apply with higher probability for larger $a$. Simulations below show that, indeed, the modified MME works better for smaller $a$. 
\end{rmk}

\begin{rmk}
As in Theorem \ref{thm:final0}, the covariance matrix is block diagonal, and therefore, asymptotically, the estimators for $(\hat{a}, \hat{b})$ and the empirical mean are independent. 
\end{rmk}

\begin{table}\scriptsize
\centering
\begin{tabular}{cc|ccc|ccc}
& & \multicolumn{3}{c|}{ $\{0.5\ln \hat{V}-\ln \hat{A}>0.5\ln (\pi/2)\}$} &  \multicolumn{3}{c}{ $\{\hat{K}>3\hat{V}\}$} \\ \hline
$a$ & $b$ & $N=10$ &  $N=20$ &  $N=50$ &  $N=10$ &  $N=20$ &  $N=50$ \\ \hline
 0.25 & 0.01 & 0.723 & 0.954 & 0.999 & 0.817 & 0.976 & 1 \\
 0.25 & 0.1 & 0.717 & 0.954 & 1 & 0.812 & 0.975 & 1 \\
 0.25 & 1 & 0.725 & 0.952 & 1 & 0.825 & 0.975 & 1 \\
 0.25 & 5 & 0.72 & 0.951 & 1 & 0.817 & 0.973 & 1 \\
 \hline
 0.5 & 0.01 & 0.559 & 0.84 & 0.99 & 0.701 & 0.922 & 0.998 \\
 0.5 & 0.1 & 0.568 & 0.847 & 0.989 & 0.718 & 0.926 & 0.998 \\
 0.5 & 1 & 0.568 & 0.848 & 0.99 & 0.708 & 0.924 & 0.998 \\
 0.5 & 5 & 0.557 & 0.849 & 0.988 & 0.704 & 0.923 & 0.999 \\
 \hline
 1 & 0.01 & 0.412 & 0.687 & 0.927 & 0.572 & 0.801 & 0.97 \\
 1 & 0.1 & 0.421 & 0.682 & 0.915 & 0.583 & 0.79 & 0.964 \\
 1 & 1 & 0.42 & 0.68 & 0.925 & 0.577 & 0.797 & 0.967 \\
 1 & 5 & 0.417 & 0.689 & 0.92 & 0.575 & 0.796 & 0.963 \\
 \hline
 2 & 0.01 & 0.318 & 0.52 & 0.77 & 0.465 & 0.639 & 0.84 \\
 2 & 0.1 & 0.317 & 0.522 & 0.773 & 0.467 & 0.64 & 0.841 \\
 2 & 1 & 0.317 & 0.518 & 0.772 & 0.474 & 0.637 & 0.838 \\
 2 & 5 & 0.314 & 0.53 & 0.775 & 0.465 & 0.646 & 0.842 \\
 \hline
 3 & 0.01 & 0.284 & 0.454 & 0.678 & 0.438 & 0.573 & 0.748 \\
 3 & 0.1 & 0.274 & 0.45 & 0.68 & 0.433 & 0.566 & 0.747 \\
 3 & 1 & 0.277 & 0.452 & 0.669 & 0.431 & 0.567 & 0.737 \\
 3 & 5 & 0.278 & 0.449 & 0.67 & 0.432 & 0.565 & 0.746 
\end{tabular}
\caption{\label{tablesupt}Simulation of the probabilities for sample size $N$ and unknown location parameter $m=0$ with respect to the existence of the estimator}
\end{table}

\subsection{Simulations} 
Simulation results for the modified estimators are given in Table \ref{table_modmom}. We find that, for all values of $a$ and $b$ considered, the MSEs of the modified MMEs $\hat{a}$ and $\hat{b}$ are reduced, often substantially, compared to the classic MME. In fact, performance is competitive even when compared to the MLE. We do, however, note that as $a$ increases, it occasionally happens that the function $L$ becomes very flat in the region close to $y=0.5\ln\hat{V} - \hat{A}(1) $, which leads to a high sensitivity of the numerical procedure used by \emph{Mathematica} with respect to small changes in the sample. This means that on rare occasions the modified MME fails completely, and gives very poor estimates. We did not exclude these rare very poor estimates from our results. 
%and these rare occasions result in an overall absurd value for the MSE, like the value $1.23\times 10^{10}$ when $(a,b)=(3,0.1)$. If we were to exclude these rare occasions in which the modified MME fails, we would see improved performance compared to the classic MME; inded, performance is better for $(a,b)=(3,0.01)$,  $(a,b)=(3,1)$ and  $(a,b)=(3,5)$. When $a=3$, we also observe some large values for the MSE of $\hat{b}$, which again results when the algorithm fails. For $a\leq2$ and $b\leq1$, we see that MSE for the modified MME $\hat{b}$ is smaller than the classic MME, although this is slightly reversed when $b=5$. 

Overall, we find that the modified MME outperforms, often substantially, classic MME in terms of smaller bias and MSE, and is quite competitive even against MLE. This suggests our modified MME could be an excellent alternative to classic MME for fitting the VG distribution to financial data (for which these parameter values are typically encountered). For other parameter values, performance is often better than for classic MME, but on rare occasions the modified MME fails completely when $a\geq3$. Therefore, as with classic MME, caution must be applied when implementing the modified MME.

\begin{center}
\begin{table}\scriptsize
\centering
\begin{tabular}{C|C|C|C|C|C|C|C}
a & b & \text{Bias}(\hat{a}) & \text{MSE}(\hat{a}) & \text{StError}(\hat{a})  & \text{Bias}(\hat{b}) & \text{MSE}(\hat{b}) & \text{StError}(\hat{b})\\ 
\hline
 0.25 & 0.01 & 4.26\text{e-3} & 7.39\text{e-4} & 8.49\text{e-4} & 1.78\text{e-5} & 4.17\text{e-6} & 6.46\text{e-5} \\
 0.25 & 0.1 & 4.07\text{e-3} & 7.17\text{e-4} & 8.37\text{e-4} & 1.22\text{e-4} & 4.02\text{e-4} & 6.34\text{e-4} \\
 0.25 & 1 & 4.36\text{e-3} & 7.19\text{e-4} & 8.37\text{e-4} & -1.19\text{e-3} & 3.95\text{e-2} & 6.28\text{e-3} \\
 0.25 & 5 & 4.12\text{e-3} & 7.23\text{e-4} & 8.4\text{e-4} & 7.69\text{e-4} & 1.01\text{e}\text{} & 3.17\text{e-2} \\
 \hline
 0.5 & 0.01 & 9.18\text{e-3} & 3.3\text{e-3} & 1.79\text{e-3} & -1.17\text{e-5} & 3.15\text{e-6} & 5.62\text{e-5} \\
 0.5 & 0.1 & 8.8\text{e-3} & 3.2\text{e-3} & 1.77\text{e-3} & -1.5\text{e-4} & 3.04\text{e-4} & 5.52\text{e-4} \\
 0.5 & 1 & 9.05\text{e-3} & 3.28\text{e-3} & 1.79\text{e-3} & -1.16\text{e-3} & 3.15\text{e-2} & 5.61\text{e-3} \\
 0.5 & 5 & 8.46\text{e-3} & 3.24\text{e-3} & 1.78\text{e-3} & 6.68\text{e-3} & 7.78\text{e-1} & 2.79\text{e-2} \\
 \hline
 1 & 0.01 & 2.64\text{e-2} & 2.17\text{e-2} & 4.59\text{e-3} & -2.48\text{e-5} & 3.33\text{e-6} & 5.77\text{e-5} \\
 1 & 0.1 & 2.71\text{e-2} & 2.18\text{e-2} & 4.59\text{e-3} & -2.88\text{e-4} & 3.28\text{e-4} & 5.72\text{e-4} \\
 1 & 1 & 2.3\text{e-2} & 2.17\text{e-2} & 4.6\text{e-3} & 2.6\text{e-3} & 3.4\text{e-2} & 5.83\text{e-3} \\
 1 & 5 & 2.29\text{e-2} & 2.21\text{e-2} & 4.64\text{e-3} & 1.19\text{e-2} & 8.55\text{e-1} & 2.92\text{e-2} \\
 \hline
 2 & 0.01 & 1.\text{e-1} & 2.39\text{e-1} & 1.51\text{e-2} & 4.4\text{e-6} & 5.42\text{e-6} & 7.36\text{e-5} \\
 2 & 0.1 & 1.01\text{e-1} & 2.35\text{e-1} & 1.5\text{e-2} & -6.7\text{e-5} & 5.3\text{e-4} & 7.28\text{e-4} \\
 2 & 1 & 1.12\text{e-1} & 2.41\text{e-1} & 1.51\text{e-2} & -4.61\text{e-3} & 5.38\text{e-2} & 7.33\text{e-3} \\
 2 & 5 & 1.07\text{e-1} & 2.41\text{e-1} & 1.52\text{e-2} & -1.04\text{e-2} & 1.36\text{e}\text{} & 3.68\text{e-2} \\
\hline
 3 & 0.01 & 3.12\text{e-1} & 1.49\text{e}\text{} & 3.73\text{e-2} & -6.17\text{e-5} & 8.37\text{e-6} & 9.15\text{e-5} \\
 3 & 0.1 & 3.62\text{e-1} & 2.58\text{e}\text{} & 4.94\text{e-2} & -1.26\text{e-3} & 8.5\text{e-4} & 9.21\text{e-4} \\
 3 & 1 & 3.11\text{e-1} & 1.53\text{e}\text{} & 3.78\text{e-2} & -4.89\text{e-3} & 8.45\text{e-2} & 9.19\text{e-3} \\
 3 & 5 & 3.07\text{e-1} & 1.55\text{e}\text{} & 3.81\text{e-2} & -1.93\text{e-2} & 2.12\text{e}\text{} & 4.6\text{e-2} 
 \end{tabular} 
\caption{\label{table_modmom} Simulation results for modified MME with known $m=0$  and sample size $N=1000$.}
\end{table}
\end{center}

%\section{The Case of 
%The General Mean the unknown location parameter}\label{secm}

%We did not consider theoretically asymptotic normality for 
%the general 
%{unknown location parameter} $m$, because this would make the analysis more complicated.
%, and superiority of the modification is clear even from the case $m = 0$. 
We also performed simulations for which the true location parameter $m=0$ is unknown. The results are reported in Table \ref{table_mumom} (classical MME) and Table \ref{table_mumodmom} (our modified MME). The results can be seen to be very similar to simulation results for known $m=0$. We remark that the MSE for classic MME for $a=3$ seem to be large in general. Therefore one needs a large number of Monte Carlo repetitions in order to determine the values more precisely and our estimates have to be treated with caution. We again stress that caution is also needed when applying modified MME for $a\geq3$, as the numerical procedure may fail. In addition, we report the estimated values of the unknown parameter $m$ in Table \ref{table_mu}. We see that the performance of this estimator is rather good, and given this good performance it is perhaps to be expected that our simulation results for the estimators $\hat{a}$ and $\hat{b}$ via classic and our modified MME are similar for the cases of known and unknown $m=0$.

\begin{center}
\begin{table}\scriptsize
\centering
\begin{tabular}{C|C|C|C|C|C|C|C}
a & b & \text{Bias}(\hat{a}) & \text{MSE}(\hat{a}) & \text{StError}(\hat{a})  & \text{Bias}(\hat{b}) & \text{MSE}(\hat{b}) & \text{StError}(\hat{b})\\ 
\hline
 0.25 & 0.01 & 4.52\text{e-2} & 9.73\text{e-3} & 2.77\text{e-3} & -3.64\text{e-4} & 2.56\text{e-5} & 1.59\text{e-4} \\
 0.25 & 0.1 & 4.46\text{e-2} & 9.74\text{e-3} & 2.78\text{e-3} & -3.69\text{e-3} & 2.5\text{e-3} & 1.58\text{e-3} \\
 0.25 & 1 & 4.47\text{e-2} & 9.74\text{e-3} & 2.78\text{e-3} & -4.24\text{e-2} & 2.19\text{e-1} & 1.48\text{e-2} \\
 0.25 & 5 & 4.39\text{e-2} & 9.56\text{e-3} & 2.76\text{e-3} & -1.93\text{e-1} & 5.65\text{e}\text{} & 7.49\text{e-2} \\
\hline
 0.5 & 0.01 & 7.17\text{e-2} & 3.14\text{e-2} & 5.12\text{e-3} & -2.99\text{e-4} & 1.63\text{e-5} & 1.27\text{e-4} \\
 0.5 & 0.1 & 6.93\text{e-2} & 3.1\text{e-2} & 5.12\text{e-3} & -2.91\text{e-3} & 1.53\text{e-3} & 1.23\text{e-3} \\
 0.5 & 1 & 7.17\text{e-2} & 3.13\text{e-2} & 5.12\text{e-3} & -3.1\text{e-2} & 1.64\text{e-1} & 1.28\text{e-2} \\
 0.5 & 5 & 6.94\text{e-2} & 3.12\text{e-2} & 5.14\text{e-3} & -1.29\text{e-1} & 4.05\text{e}\text{} & 6.35\text{e-2} \\
\hline
 1 & 0.01 & 1.33\text{e-1} & 1.3\text{e-1} & 1.06\text{e-2} & -2.41\text{e-4} & 1.39\text{e-5} & 1.18\text{e-4} \\
 1 & 0.1 & 1.35\text{e-1} & 1.31\text{e-1} & 1.06\text{e-2} & -2.71\text{e-3} & 1.33\text{e-3} & 1.15\text{e-3} \\
 1 & 1 & 1.26\text{e-1} & 1.26\text{e-1} & 1.05\text{e-2} & -1.58\text{e-2} & 1.46\text{e-1} & 1.21\text{e-2} \\
 1 & 5 & 1.27\text{e-1} & 1.28\text{e-1} & 1.06\text{e-2} & -8.04\text{e-2} & 3.56\text{e}\text{} & 5.96\text{e-2} \\
\hline
 2 & 0.01 & 3.03\text{e-1} & 7.93\text{e-1} & 2.65\text{e-2} & -1.6\text{e-4} & 1.46\text{e-5} & 1.21\text{e-4} \\
 2 & 0.1 & 3.14\text{e-1} & 8.11\text{e-1} & 2.67\text{e-2} & -1.68\text{e-3} & 1.53\text{e-3} & 1.24\text{e-3} \\
 2 & 1 & 3.27\text{e-1} & 8.15\text{e-1} & 2.66\text{e-2} & -2.42\text{e-2} & 1.47\text{e-1} & 1.21\text{e-2} \\
 2 & 5 & 3.14\text{e-1} & 8.3\text{e-1} & 2.7\text{e-2} & -6.96\text{e-2} & 4.05\text{e}\text{} & 6.36\text{e-2} \\
\hline
 3 & 0.01 & 6.63\text{e-1} & 4.67\text{e}\text{} & 6.5\text{e-2} & -2.46\text{e-4} & 1.75\text{e-5} & 1.32\text{e-4} \\
 3 & 0.1 & 7.43\text{e-1} & 6.54\text{e}\text{} & 7.74\text{e-2} & -3.1\text{e-3} & 1.74\text{e-3} & 1.32\text{e-3} \\
 3 & 1 & 6.78\text{e-1} & 4.15\text{e}\text{} & 6.07\text{e-2} & -2.02\text{e-2} & 1.87\text{e-1} & 1.37\text{e-2} \\
 3 & 5 & 1.57\text{e}\text{} & 8.13\text{e}3 & 2.85\text{e}\text{} & -1.21\text{e-1} & 4.47\text{e}\text{} & 6.68\text{e-2}
 \end{tabular}
\caption{\label{table_mumom} Simulation results for classic MME with unknown $m=0$  and sample size $N=1000$.}
\end{table}
\end{center}

\begin{center}
\begin{table}\scriptsize
\centering
\begin{tabular}{C|C|C|C|C|C|C|C}
a & b & \text{Bias}(\hat{a}) & \text{MSE}(\hat{a}) & \text{StError}(\hat{a})  & \text{Bias}(\hat{b}) & \text{MSE}(\hat{b}) & \text{StError}(\hat{b})\\ 
\hline
 0.25 & 0.01 & 1.08\text{e-2} & 9.13\text{e-4} & 8.93\text{e-4} & -2.37\text{e-4} & 4.06\text{e-6} & 6.33\text{e-5} \\
 0.25 & 0.1 & 1.05\text{e-2} & 8.89\text{e-4} & 8.83\text{e-4} & -2.4\text{e-3} & 3.91\text{e-4} & 6.21\text{e-4} \\
 0.25 & 1 & 1.08\text{e-2} & 8.98\text{e-4} & 8.84\text{e-4} & -2.61\text{e-2} & 3.88\text{e-2} & 6.18\text{e-3} \\
 0.25 & 5 & 1.06\text{e-2} & 8.96\text{e-4} & 8.85\text{e-4} & -1.26\text{e-1} & 9.87\text{e-1} & 3.12\text{e-2} \\
\hline
 0.5 & 0.01 & 1.41\text{e-2} & 3.51\text{e-3} & 1.82\text{e-3} & -1.16\text{e-4} & 3.12\text{e-6} & 5.57\text{e-5} \\
 0.5 & 0.1 & 1.37\text{e-2} & 3.42\text{e-3} & 1.8\text{e-3} & -1.19\text{e-3} & 3.01\text{e-4} & 5.47\text{e-4} \\
 0.5 & 1 & 1.41\text{e-2} & 3.51\text{e-3} & 1.82\text{e-3} & -1.18\text{e-2} & 3.11\text{e-2} & 5.57\text{e-3} \\
 0.5 & 5 & 1.36\text{e-2} & 3.45\text{e-3} & 1.81\text{e-3} & -4.75\text{e-2} & 7.65\text{e-1} & 2.76\text{e-2} \\
 \hline
 1 & 0.01 & 3.06\text{e-2} & 2.22\text{e-2} & 4.62\text{e-3} & -7.45\text{e-5} & 3.31\text{e-6} & 5.75\text{e-5} \\
 1 & 0.1 & 3.15\text{e-2} & 2.23\text{e-2} & 4.62\text{e-3} & -8.04\text{e-4} & 3.26\text{e-4} & 5.7\text{e-4} \\
 1 & 1 & 2.74\text{e-2} & 2.22\text{e-2} & 4.63\text{e-3} & -2.63\text{e-3} & 3.38\text{e-2} & 5.81\text{e-3} \\
 1 & 5 & 2.74\text{e-2} & 2.26\text{e-2} & 4.67\text{e-3} & -1.43\text{e-2} & 8.49\text{e-1} & 2.91\text{e-2} \\
 \hline
 2 & 0.01 & 1.06\text{e-1} & 2.43\text{e-1} & 1.52\text{e-2} & -3.33\text{e-5} & 5.4\text{e-6} & 7.35\text{e-5} \\
 2 & 0.1 & 1.08\text{e-1} & 2.38\text{e-1} & 1.5\text{e-2} & -4.55\text{e-4} & 5.28\text{e-4} & 7.26\text{e-4} \\
 2 & 1 & 1.18\text{e-1} & 2.44\text{e-1} & 1.52\text{e-2} & -8.41\text{e-3} & 5.35\text{e-2} & 7.31\text{e-3} \\
 2 & 5 & 1.14\text{e-1} & 2.45\text{e-1} & 1.52\text{e-2} & -3.02\text{e-2} & 1.35\text{e}\text{} & 3.67\text{e-2} \\
\hline
 3 & 0.01 & 3.23\text{e-1} & 1.53\text{e}\text{} & 3.78\text{e-2} & -1.01\text{e-4} & 8.32\text{e-6} & 9.12\text{e-5} \\
 3 & 0.1 & 3.67\text{e-1} & 2.27\text{e}\text{} & 4.62\text{e-2} & -1.6\text{e-3} & 8.47\text{e-4} & 9.19\text{e-4} \\
 3 & 1 & 3.21\text{e-1} & 1.55\text{e}\text{} & 3.81\text{e-2} & -8.49\text{e-3} & 8.41\text{e-2} & 9.17\text{e-3} \\
 3 & 5 & 3.14\text{e-1} & 1.54\text{e}\text{} & 3.8\text{e-2} & -3.61\text{e-2} & 2.1\text{e}\text{} & 4.59\text{e-2}
 \end{tabular}
\caption{\label{table_mumodmom} Simulation results for modified MME with unknown $m=0$  and sample size $N=1000$.}
\end{table} 
\end{center}

\begin{center}
\begin{table}{\scriptsize
\begin{tabular}{C|C|C|C|C}
 a & b & \text{Bias}(\hat{\mu}) & \text{MSE}(\hat{\mu}) & \text{StError}(\hat{\mu}) \\ 
 \hline
 0.25 & 0.01 & 1.31\text{e-5} & 2.55\text{e-6} & 5.05\text{e-5} \\
 0.25 & 0.1 & 5.95\text{e-5} & 2.49\text{e-5} & 1.58\text{e-4} \\
 0.25 & 1 & -1.93\text{e-5} & 2.48\text{e-4} & 4.98\text{e-4} \\
 0.25 & 5 & -7.3\text{e-4} & 1.23\text{e-3} & 1.11\text{e-3} \\
  \hline
 0.5 & 0.01 & 2.96\text{e-5} & 5.05\text{e-6} & 7.1\text{e-5} \\
 0.5 & 0.1 & -1.24\text{e-4} & 4.96\text{e-5} & 2.23\text{e-4} \\
 0.5 & 1 & 2.99\text{e-4} & 4.98\text{e-4} & 7.06\text{e-4} \\
 0.5 & 5 & 4.47\text{e-4} & 2.48\text{e-3} & 1.58\text{e-3} \\
  \hline
 1 & 0.01 & 1.58\text{e-5} & 1.01\text{e-5} & 1.\text{e-4} \\
 1 & 0.1 & -7.13\text{e-5} & 9.89\text{e-5} & 3.14\text{e-4} \\
 1 & 1 & -5.4\text{e-4} & 9.76\text{e-4} & 9.88\text{e-4} \\
 1 & 5 & 8.24\text{e-4} & 4.99\text{e-3} & 2.23\text{e-3} \\
  \hline
 2 & 0.01 & 1.02\text{e-5} & 2.01\text{e-5} & 1.42\text{e-4} \\
 2 & 0.1 & 1.66\text{e-4} & 2.01\text{e-4} & 4.48\text{e-4} \\
 2 & 1 & -1.71\text{e-4} & 2.03\text{e-3} & 1.42\text{e-3} \\
 2 & 5 & 4.32\text{e-4} & 1.01\text{e-2} & 3.18\text{e-3} \\
  \hline
 3 & 0.01 & -4.54\text{e-5} & 2.98\text{e-5} & 1.72\text{e-4} \\
 3 & 0.1 & -3.7\text{e-4} & 2.99\text{e-4} & 5.47\text{e-4} \\
 3 & 1 & 3.69\text{e-4} & 3.02\text{e-3} & 1.74\text{e-3} \\
 3 & 5 & -2.63\text{e-3} & 1.47\text{e-2} & 3.83\text{e-3} \\
\end{tabular}}
\caption{\label{table_mu} Simulation results for estimation of unknown $m=0$ with classic MME  and sample size $N=1000$.}
\end{table}
\end{center}

\section{Conclusion and Further Research}\label{sec4} 

We tested the classic MME for a symmetric variance-gamma (generalized symmetric Laplace) distribution. Using simulations and the delta method, we showed that caution must be used in implementing this method in practice as performance is often not acceptable. This runs contrary to some remarks in the existing literature, some of which was cited in the Introduction, that may give the impression that MME works across the full range of parameter values for the variance-gamma distribution (and time series models based on it). 

However, we also produced positive results. We modified MME for symmetric variance-gamma by switching to the first two absolute moments (mean absolute deviation) instead of regular ones. Our simulation results showed that our modified estimator is more efficient than for the classic MME across a broad range of parameter values, in particular those encountered in financial modelling. However, like classic MME, caution must be used in implementing this modified MME as when $a\geq3$ performance degrades. Our study suggests using absolute centered moments for other symmetric distributions offers a possibility for improved performance. 

A natural question is how to extend this modified MME to the asymmetric variance-gamma (generalized asymmetric Laplace)\gr{.}
% and, more generally, Gaussian-Laplace distributions. T
%This presents a problem: Increasing the number of parameters further increases asymptotic variances. If the classic or modified MME theoretically works, in the sense that it is consistent and asymptotically normal, but limiting variances are very high, then it is of limited practical use. In addition, 
It is, however, not so easy to solve the system of equations for these cases, at least not nearly as easily as for symmetric variance-gamma. \gr{Indeed, for the asymmetric Laplace distribution~\eqref{eq:mixture} with $c\not=0$, formulas for the absolute raw moments $\mathbb{E}[|X|^k]$ take a complicated form involving the hypergeometric function (see \cite[Theorem 2.1]{gaunt23}) and, to the best knowledge of the authors, explicit formulas are not available in the literature for the absolute centered raw moments $\mathbb E[|X - (m-ac)|^k]$.}

\begin{comment}
For example, consider the generalized asymmetric Laplace distribution~\eqref{eq:mixture}. Then 
\[\mathbb E[|X - \gr{(m-ac)}|]=\left[\frac{2b}{\pi}\right]^{1/2}\frac{\Gamma(a + 0.5)}{\Gamma(a)}, \quad \mathrm{Var}(X)= \frac{c^2}{a^2} + \frac{b}{a}.\]
%$$
%\mathbb E[|X - m|] = \mathbb E\left[\mathbb E[|X - m|\mid G]\right] = \mathbb E[\sqrt{2bG/\pi}] = \left[\frac{2b}{\pi}\right]^{1/2}\frac{\Gamma(a + 0.5)}{\Gamma(a)}. 
%$$
%Next, compute the variance: 
%$$
%\mathrm{Var}(X) = \mathrm{Var}(\mathbb E(X\mid G)) + \mathbb E(\mathrm{Var}(X\mid G)) = \mathrm{Var}(m + cG) + \mathbb E(bG) = \frac{c^2}{a^2} + \frac{b}{a}.
%$$
(see \cite{vgsurvey}). We need a third equation. For this purpose, we can take, for example, the third absolute moment. But 
%as we see, 
it is not clear how to solve this system of equations. Still, we cannot reject the hope that some further modification of the MME could work in these cases. We leave this topic for future research.
\end{comment}

\appendix

\section{Proofs}

\label{appendix}

\noindent {\emph{Proof of Theorem~\ref{thm1}.}} Just solve for $a$ and $b$ this system of two equations~\eqref{eq:moments}:
\begin{equation}
\label{eq:solution}
a = \frac{3V^2}{K-3V^2},\quad b = \frac{K}{3V} - V.
\end{equation}
Next, plug in $\hat{V}$ instead of $V$, and $\hat{K}$ instead of $K$. By the Strong Law of Large Numbers, $\overline{X}\to m$ $\hat{V} \to V$ a.s. and $\hat{K} \to K$ a.s. as $N \to \infty$. Next, $\Phi : (K, V,m) \mapsto (a, b,m)$ in~\eqref{eq:solution} is a continuous function. Thus we get almost surely $(\hat{a}, \hat{b},\hat{m}) = \Phi(\hat{K}, \hat{V},\overline{X}) \to (a, b,m)$. 
%This proves consistency. 
\hfill $\Box$

\vspace{3mm}

\noindent {\emph{Proof of Lemma~\ref{lemkv}.}} We consider the (open) feasibility set 
\begin{equation}
\label{eq:O-classic}
\mathcal O := \{(v, k) \in \mathbb R^2\mid k > 3v^2,\, v > 0\}.
\end{equation} 
Therefore, there exists an $\varepsilon$-neighborhood $\mathcal U \subseteq \mathcal O$ of $(V, K)$. It suffices to show the estimate
$$
\mathbb P((\hat{V}', \hat{K}') \in \mathcal U) \ge 1 - \frac CN.
$$
But this follows from the estimate (where $X$ has the SVG distribution):
\begin{equation}
\label{eq:exp-est0}
M := \mathbb E\left[(\hat{V}' - V)^2 + (\hat{K}' - K)^2\right] = \mathrm{Var}(\hat{V}') + \mathrm{Var}(\hat{K}') = \frac{\mathrm{Var}(X^2)}N + \frac{\mathrm{Var}(X^4)}N = \frac CN,
\end{equation}
and Markov's inequality (with the $M$ defined in~\eqref{eq:exp-est0}):
$$
\mathbb P((\hat{V}' - V)^2 + (\hat{K}' - K)^2 \ge \varepsilon^2) \le \frac{M}{\varepsilon^2}.
$$ \hfill $\Box$

\vspace{3mm}

\noindent{\emph{Proof of Theorem \ref{thm:original}.}} State the Central Limit Theorem for~\eqref{eq:modifications}:
$$
\sqrt{N}\left[(\hat{V}, \hat{K}) - (V, K)\right] = \sqrt{N}\left[\sum\limits_{i=1}^N(X_i^2, X_i^4) - (V, K)\right] \to_d \mathcal N_2([0, 0], A),
$$
since $(X_i^2, X_i^4)$ are i.i.d. random vectors with finite covariance matrix. Now, let us find the limiting covariance matrix. We have that
\begin{align}
\label{eq:A}
\begin{split}
C &= 
\begin{bmatrix}
\mathrm{Var}(X^2) & \mathrm{Cov}(X^2, X^4) \\  \mathrm{Cov}(X^2, X^4) & \mathrm{Var}(X^4)
\end{bmatrix} \\ & = 
\begin{bmatrix}
\mathbb E[X^4] - (\mathbb E[X^2])^2 & \mathbb E[X^6] - \mathbb E[X^2]\cdot \mathbb E[X^4] \\
\mathbb E[X^6] - \mathbb E[X^2]\cdot \mathbb E[X^4] & \mathbb E[X^8] - (\mathbb E[X^4])^2
\end{bmatrix}.
\end{split}
\end{align}
Using the representation (\ref{vgrep}) and standard formulas for moments of gamma and normal random variables, or appealing to \cite[Proposition 4.1.6]{Book}, we have the following formulas for  the moments of orders 6 and 8:
\begin{equation}\label{eq:6-8}
\mathbb E[X^6]= 15a(a+1)(a+2)b^3 , 
\quad \mathbb E[X^8] =105a(a+1)(a+2)(a+3)b^4.
\end{equation}
\begin{comment}
Let us find the higher-order moments. First, $\mathbb E[Z^6] = 15$ and $\mathbb E[Z^8] = 105$ for $Z \sim \mathcal N(0, 1)$. Next, we recall the standard formulas for the third and fourth moments of the gamma distribution: $\mathbb E[G^3]=a(a+1)(a+2)$ and $\mathbb E[G^4]=a(a+1)(a+2)(a+3)$. Thus we compute the moments of orders 6 and 8:
\begin{align}
\label{eq:6-8}
\begin{split}
\mathbb E[X^6] &= \mathbb E[b^3G^3Z^6] = b^3\cdot \mathbb E[G^3] \cdot \mathbb E[Z^6] = 15a(a+1)(a+2)b^3,\\
\mathbb E[X^8] &= \mathbb E[b^4G^4Z^8] = b^4\cdot \mathbb E[G^4] \cdot \mathbb E[Z^8] = 105a(a+1)(a+2)(a+3)b^4.
\end{split}
\end{align}
\end{comment}
Plug these moments~\eqref{eq:6-8}, together with $V$ and $K$ from~\eqref{eq:moments}, into the matrix from~\eqref{eq:A}:
\begin{align*}
C &= 
\begin{bmatrix}
3a(a+1)b^2 - a^2b^2 & 15a(a+1)(a+2)b^3 - 3a^2(a+1)b^3\\ 
15a(a+1)(a+2)b^3 - 3a^2(a+1)b^3 & 105a(a+1)(a+2)(a+3)b^4 - 9a^2(a+1)^2b^4
\end{bmatrix}
\\ & = 
\begin{bmatrix}
(2a^2 + 3a)b^2 & a(a+1)(12a + 30)b^3 \\ a(a+1)(12a + 30)b^3 & a(a+1)(96a^2 + 516a + 630)b^4
\end{bmatrix}.
\end{align*}
Finally, let us compute the Jacobian of the function $\Phi$ from~\eqref{eq:solution}:
\begin{equation}
\label{eq:Jacobian}
J = \begin{bmatrix}
\frac{\partial a}{\partial V} & \frac{\partial a}{\partial K} \\ 
\frac{\partial b}{\partial V} & \frac{\partial b}{\partial K}
\end{bmatrix}
= \begin{bmatrix}
\frac{6KV}{(K-3V^2)^2} & -\frac{3V^2}{(K-3V^2)^2} \\
-\frac{K}{3V^2} - 1 & \frac1{3V}
\end{bmatrix}.
\end{equation}
Plugging $V$ and $K$ from~\eqref{eq:moments} into~\eqref{eq:Jacobian}, we get:
$$
J = \begin{bmatrix}
\frac{2(a+1)}{b^2} & -\frac1{3b^4} \\ -2-\frac1a & \frac1{3ab^2} 
\end{bmatrix}.
$$
Applying the bivariate $\delta$ method to $\Phi$, as in \cite[Section 5.5]{Textbook}, we get the CLT for $(\hat{a}, \hat{b})$ with limiting covariance matrix $\Sigma = JCJ^T$. \hfill $\Box$

\vspace{3mm}

\noindent{\emph{Proof of Theorem~\ref{thm:final0}.}} Assume without loss of generality that $m = 0$. There are two facts:
\begin{equation}
\label{eq:close-10}
\sqrt{N}(\hat{V}' - \hat{V}) \rightarrow_d 0,\, N \to \infty;
\end{equation}
\begin{equation}
\label{eq:close-20}
\sqrt{N}(\hat{K}' - \hat{K}) \rightarrow_d 0,\, N \to \infty.
\end{equation}
Assuming we proved~\eqref{eq:close-10} and~\eqref{eq:close-20}, let us complete the proof of Theorem~\ref{thm:final0}. Similarly to the proof of Theorem~\ref{thm:original}, we can get
\begin{equation}
\label{eq:CLT-30}
\sqrt{N}[(\overline{X}, \hat{V}, \hat{K}) - (m, V, K)] \to_d \mathcal N_3(0, C_3),
\end{equation}
where the $3\times 3$ limiting covariance matrix $C_3$ is given by 
$$
C_3 = 
\begin{bmatrix}
\mathbb{E}[X^2] & \mathbb{E}[X^3] & \mathbb{E}[X^5] \\
\mathbb{E}[X^3] & \mathbb{E}[X^4] & \mathbb{E}[X^6] \\
\mathbb{E}[X^5] & \mathbb{E}[X^6] & \mathbb{E}[X^8]
\end{bmatrix}.
$$
Its 12 and 13 elements are odd moments. By symmetry of $X$, these are equal to zero. Therefore, the matrix $C_3$ is block diagonal. Its $2\times 2$ block is 
\begin{equation}
\label{eq:block-C0}
C_3 = \mathrm{diag}(ab, C_2).
\end{equation}
Next, apply the Slutsky theorem to~\eqref{eq:CLT-3}. Together with~\eqref{eq:close-10} and~\eqref{eq:close-20}, we conclude that the statement of~\eqref{eq:CLT-30} holds even if we replace $\hat{K}$ with $\hat{K}'$ and $\hat{V}$ with $\hat{V}'$. Next, apply the delta method as before to $(\hat{V}', \hat{K}')$; or, rather, to the entire $(\overline{X}, \hat{V}', \hat{K}')$ with the mapping $\Psi_3(w, x, y) := (w, \Psi(x, y))$. The Jacobi matrix $J_3$ is also block-diagonal: 
\begin{equation}
\label{eq:block-J0}
J_3 = \mathrm{diag}(1, J_2).
\end{equation}
Combining~\eqref{eq:block-C0} and~\eqref{eq:block-J0}, we conclude that the limiting covariance matrix $\Sigma_3$ is block-diagonal as well: $\Sigma_3 = J_3C_3J^T_3 = \mathrm{diag}(ab, \Sigma_2)$. This concludes the proof of Theorem~\ref{thm:final0}.

It remains only to show \eqref{eq:close-10} and \eqref{eq:close-20}. First, \eqref{eq:close-10} follows from direct computation: $\sqrt{N}(\hat{V}' - \hat{V}) = -\sqrt{N}|\overline{X}|^2$, and the CLT: $\sqrt{N}\overline{X} \to_d Z \sim \mathcal N(0, \sigma^2)$ as $N \to \infty$. Combining these observations, we get that the left-hand side of~\eqref{eq:close-10} behaves asymptotically as $-\sqrt{N}(Z/\sqrt{N})^2 = -Z^2/\sqrt{N} \to 0$. Similarly, \eqref{eq:close-20} follows from expanding $\hat{K}$, we have: 
\begin{equation}
    \label{eq:empirical-kurt}
\hat{K}' - \hat{K} = -4\overline{X^3}\cdot\overline{X} + 6\overline{X^2}\cdot\overline{X}^2 - 4\overline{X}^4,
\end{equation}
where we define the empirical average of the $j$th moment:
\begin{equation}
\label{eq:average-moments}
\overline{X^j} = \frac1N\sum\limits_{l=1}^NX^j_l. 
\end{equation}
Applying the CLT to \eqref{eq:average-moments} for $j = 3$, and using the symmetry of $X$ around $0$, we get:
\begin{equation}
    \label{eq:CLT-3}
\sqrt{N}\overline{X^3} \rightarrow_d Z \sim \mathcal N(0, \mathrm{Var}(X^3)),\quad N \to \infty.
\end{equation}
By the Strong Law of Large Numbers applied to \eqref{eq:average-moments}, we have:
\begin{equation}
\label{eq:SLLN-moments}
\overline{X^j} \to \mathbb E[X^j]\quad \mbox{almost surely}\quad N \to \infty.
\end{equation}
Combining~\eqref{eq:empirical-kurt},~\eqref{eq:CLT-3}, \eqref{eq:SLLN-moments} and applying the Slutsky theorem, we prove~\eqref{eq:close-20}. \hfill $\Box$

\vspace{3mm}

\noindent{\emph{Proof of Lemma~\ref{lemma:decreasing}}.} {\it Step 1.} Let us show that $L'(x) < 0$ for all $x > 0$. Consider the {\it digamma} function $\psi(x) := (\ln\Gamma(x))' = \Gamma'(x)/\Gamma(x)$. We have that $L'(x) = \frac{0.5}{x} + \psi'(x) - \psi'(x + 0.5)$. Applying the two-sided inequality 
$
\ln x - \frac1x < \psi(x) < \ln x - \frac{1}{2x},
$
$x>0$ (see \cite{digamma}) we get the upper estimate: 
\begin{equation}
\label{eq:main-estimate}
L'(x) \le \frac1{2x} + \ln x - \frac1{2x} - \left(\ln(x + 0.5) - \frac1{x + 0.5}\right) = \ln\frac{x}{x + 0.5} + \frac1{x + 0.5} =: M(x).
\end{equation}
The derivative of $M$ is $M'(x) = [x(2x+1)^2]^{-1}$ thus $M$ is increasing. Next, $M(x) \to 0$ as $x \to \infty$. Thus $M(x) < M(\infty) = 0$ for all $x > 0$. Combining this observation with~\eqref{eq:main-estimate}, we complete the proof that $L' < 0$.

\smallskip

{\it Step 2.} Let us show that $L(0+) = \infty$: It follows from convergence 
$$
0.5\ln x + \ln\Gamma(x) = 0.5\ln x + \ln\frac{\Gamma(x+1)}{x} = \ln\Gamma(x+1) - 0.5\ln x \to \ln 1 - 0.5\ln 0 = +\infty
$$
and $\ln\Gamma(x+0.5) \to \ln\Gamma(0.5)$ as $x \to 0+$. 

\smallskip

{\it Step 3.} Finally, let us show $L(\infty) = 0.5\ln(0.5\pi)$. Indeed, as $x \to \infty$, we have
\begin{align*}
L(x) + L(x+0.5) - 2\cdot 0.5\ln(\pi/2) &= 0.5\ln x + 0.5\ln(x + 0.5) + \ln \Gamma(x) - \ln \Gamma(x+1) \\ & = 0.5\ln x + 0.5\ln(x + 0.5) - \ln x = 0.5\ln((x+0.5)/x) \to 0.
\end{align*}
So $2L(\infty) - 2\cdot 0.5\ln(\pi/2) = 0$, thus $L(\infty) = 0.5\ln(\pi/2)$. \hfill $\Box$

\vspace{3mm}

\noindent{\emph{Proof of Theorem \ref{normod}}.} We use the same techniques as in the proof of Theorem~\ref{thm:original}. Namely, we use continuity and smoothness of the mapping 
$$
\Psi : (x, y) \mapsto \left[\ell(0.5\ln y - \ln x),\, \frac{y}{\ell(0.5\ln y - \ln x)}\right]
$$
which maps $(A, V)$ into $(a, b)$ and similarly $(\hat{A}, \hat{V})$ into $(\hat{a}, \hat{b})$. By the Central Limit Theorem, 
\begin{equation}
\label{eq:classic}
\sqrt{N}\left[(\hat{A}, \hat{V}) - (A, V)\right] \rightarrow_d \mathcal N_2\left(\begin{bmatrix} 0 \\ 0\end{bmatrix},  C_2=\begin{bmatrix} V & T \\ T & K \end{bmatrix}\right)
\end{equation}
where $T = \mathbb E[|X|^3]$ 
%is the absolute centered third moment, 
and $K = \mathbb E[|X|^4]$.
%is the fourth central moment, defined in~\eqref{eq:moments}. 
Applying the $\delta$ method, we complete the proof. 
%This also gives us a way to find the limiting covariance matrix $\Sigma_2$. 
\hfill $\Box$

\vspace{3mm}

\noindent{\emph{Proof of Lemma~\ref{lemma:applicability-estimate}}.} This is very similar to the proof of Lemma~\ref{lemkv}. The following set is open:
\begin{equation}
\label{eq:O}
\mathcal O := \{(a, v) \in \mathbb R^2\mid a, v > 0;\quad 0.5\ln v - \ln a > 0.5\ln(\pi/2)\}.
\end{equation}
Therefore, there exists an $\varepsilon$-neighborhood $\mathcal U \subseteq \mathcal O$ of $(A, V)$. It suffices to show the estimate
$$
\mathbb P((\hat{A}', \hat{V}') \in \mathcal U) \ge 1 - \frac CN.
$$
But this follows from the estimate (where $X$ has the SVG distribution):
\begin{equation}
\label{eq:exp-est}
M := \mathbb E\left[(\hat{A}' - A)^2 + (\hat{V}' - V)^2\right] = \mathrm{Var}(\hat{A}') + \mathrm{Var}(\hat{V}') = \frac{\mathrm{Var}(|X|)}N + \frac{\mathrm{Var}(X^2)}N = \frac CN
\end{equation}
and the Markov inequality (with the $M$ defined in~\eqref{eq:exp-est}):
$$
\mathbb P((\hat{A}' - A)^2 + (\hat{V}' - V)^2 \ge \varepsilon^2) \le \frac{M}{\varepsilon^2}.
$$
\hfill $\Box$

\vspace{3mm}

\noindent{\emph{Proof of Theorem \ref{thm:final}}.} This is very similar to the proof of Theorem \ref{thm:final0}. Without loss of generality we assume $m = 0$. There are two facts:
\begin{equation}
\label{eq:close-1}
\sqrt{N}(\hat{A}' - \hat{A}) \rightarrow_d 0,\, N \to \infty,
\end{equation}
and~\eqref{eq:close-10}, shown in the proof of Theorem \ref{thm:final0}. Assuming we proved~\eqref{eq:close-1} and~\eqref{eq:close-10}, let us complete the proof of Theorem~\ref{thm:final}. Similarly to the proof of Theorem~\ref{thm:original}, we can get
\begin{equation}
\label{eq:CLT-3D}
\sqrt{N}[(\overline{X}, \hat{A}, \hat{V}) - (m, A, V)] \to_d \mathcal N_3(0, C_3),
\end{equation}
where the $3\times 3$ limiting covariance matrix $C_3$ is given by 
$$
C_3=\begin{bmatrix}
\mathbb E[X^2] & \mathbb E[X|X|] & \mathbb E[X^3] \\
\mathbb E[X|X|] & \mathbb E[|X|^2] & \mathbb E[|X|X^2] \\
\mathbb E[X^3] & \mathbb E[|X|X^2] & \mathbb E[X^4]
\end{bmatrix}.
$$
By symmetry of the distribution of $X$ and the fact that $\mathbb E[X] = 0$, we get: $\mathbb E[X|X|] = \mathbb E[X^3] = 0$. Therefore, the matrix $C_3$ is block diagonal. Its $2\times 2$ block is the same as the limiting covariance matrix in~\eqref{eq:classic}: 
\begin{equation}
\label{eq:block-C}
C_3 = \mathrm{diag}(ab, C_2).
\end{equation}
Next, apply the Slutsky theorem to~\eqref{eq:CLT-3D}. Together with~\eqref{eq:close-1} and~\eqref{eq:close-10}, we conclude that the statement of~\eqref{eq:CLT-3D} holds even if we replace $\hat{A}$ with $\hat{A}'$ and $\hat{V}$ with $\hat{V}'$. Next, apply the delta method as before to $(\hat{A}', \hat{V}')$; or, rather, to the entire $(\overline{X}, \hat{A}', \hat{V}')$ with the mapping $\Psi_3(w, x, y) := (w, \Psi(x, y))$. The Jacobi matrix $J_3$ is also block-diagonal: 
\begin{equation}
\label{eq:block-J}
J_3 = \mathrm{diag}(1, J_2).
\end{equation}
Combining~\eqref{eq:block-C} and~\eqref{eq:block-J}, we conclude that the limiting covariance matrix $\Sigma_3$ is block-diagonal as well: 
$\Sigma_3 = J_3C_3J^T_3 = \mathrm{diag}(ab, \Sigma_2)$. This concludes the proof of Theorem~\ref{thm:final}.

It remains only to show~\eqref{eq:close-1}. The first statement follows from \cite[Theorem 2.2]{BabuRao}. We again apply the simple observation that the SVG distribution is symmetric: its mean and median coincide. In addition, the value of the density function at the mean is strictly positive. $\hfill \Box$

\vspace{3mm}

\noindent{\emph{Proof of Lemma~\ref{lemma:feasibility-estimate}}.} Without loss of generality, assume $m = 0$. Consider the $p$-norm 
$$
|\mathbf{x}|_p := \left[\frac1N\sum\limits_{k=1}^N|x_i|^p\right]^{1/p},\quad \mathbf{x} = (x_1, \ldots, x_N) \in \mathbb R^N.
$$
The vector $\mathbf{e} := (1, \ldots, 1)$ has $|\mathbf{e}|_p = 1$. For the SVG sample $X_1, \ldots, X_N$, define the vector $\mathbf{X} = (X_1, \ldots, X_N)$. By the Minkowski inequality, 
\begin{equation}
\label{eq:minkowski}
\left||\mathbf{X} - \mathbf{e}\overline{X}|_p - |\mathbf{X}|_p \right| \le |\mathbf{e}\overline{X}|_p = |\overline{X}|.
\end{equation}
Applying~\eqref{eq:minkowski} to $p = 1$ and $p = 2$, we get: 
\begin{equation}
\label{eq:basic-estimates}
|\hat{A}' - \hat{A}| \le |\overline{X}|,\quad \left|\sqrt{\hat{V}'} - \sqrt{\hat{V}}\right| \le |\overline{X}|. 
\end{equation}
Recall the notation~\eqref{eq:O} from the proof of Lemma~\ref{lemma:applicability-estimate}. From~\eqref{eq:basic-estimates}, there exist $\delta > 0$ and $\varepsilon' > 0$ such that if $|\overline{X}| < \delta$ and $(\hat{A}', \hat{V}')$ is in the $\varepsilon'$-neighbourhood $\mathcal U'$ of $(A, V)$, then $(\hat{A}, \hat{V}) \in \mathcal O$, and therefore the modified MME is applicable. We already have an estimate of the type (from the proof of Lemma~\ref{lemma:applicability-estimate}):
\begin{equation}
\label{eq:new-est-neighborhood}
\mathbb P((\hat{A}', \hat{V}') \in \mathcal U') \ge 1 - \frac{C'}N.
\end{equation}
We can get another estimate
\begin{equation}
\label{eq:Mkv}
\mathbb P(|\overline{X}| \ge \delta) \le \frac{\mathbb E[|\overline{X}|^2]]}{\delta^2} = \frac{\mathbb E[X^2]}{N\delta^2} = \frac{ab}{N\delta^2}. 
\end{equation}
From~\eqref{eq:new-est-neighborhood} and~\eqref{eq:Mkv} together, we can get the required estimate. \hfill $\Box$

\section*{Acknowledgements}
We would like to thank the two reviewers and the AE for their constructive comments and suggestions that have lead to a much improved article.  AF is funded in part by ARC Consolidator grant from ULB and FNRS Grant CDR/OL J.0197.20.

\bibliographystyle{plain}

\end{document}